\documentstyle[a4,11pt,mssymb,epsf]{article}
\newcommand{\indrm}[1]{{\mbox{\scriptsize #1}}}
\newcommand{\indfrm}[1]{{\mbox{\tiny #1}}}
\newcommand{\murf}{\mu_\indfrm{R}}
\newcommand{\mur}{\mu_\indrm{R}}
\newcommand{\dtrf}{\Delta t_\indfrm{R}}
\newcommand{\dtr}{\Delta t_\indrm{R}}
\newcommand{\sig}[1]{\sigma[#1]}
\newcommand{\sigs}[1]{\sigma^2[#1]}
\catcode`\@=11
\long\def\@makecaption#1#2{%
   \vskip 10\p@
   \setbox\@tempboxa\hbox{{\bf\footnotesize #1:} #2}%
   \ifdim \wd\@tempboxa >\hsize
       {{\bf\footnotesize #1:} #2}\par
     \else
       \hbox to\hsize{\hfil\box\@tempboxa\hfil}%
   \fi}
\catcode`\@=12

\begin{document}
\noindent
{\LARGE\bf
The lightcurve reconstruction method for measuring the time delay of
gravitational lens systems 
}

\vspace{5mm}
{\Large
\centerline{Bernhard Geiger \& Peter Schneider}
}
\vspace{2mm}
\centerline{Max-Planck-Institut f\"ur Astrophysik}
\centerline{Karl-Schwarzschild-Str. 1, Postfach 1523}
\centerline{D-85740 Garching bei M\"unchen, Germany}
\vspace{5mm}

\smallskip
%
% abstract
%
\noindent
{\large\bf Abstract:}
We propose a new technique to measure the time delay of radio-loud
gravitational lens systems, which does not rely on the excessive use of
interferometric observations. Instead, the method is based on single-dish 
flux density monitoring of the (unresolved) lens system's total lightcurve,
combined with additional interferometric measurements of the flux
density ratio at a few epochs during that monitoring period.

The basic idea of the method is to reconstruct the individual
image lightcurves from the observed total lightcurve by assuming a
range of potential values for the time delay and the magnification
ratio of the images. It is then possible to single out the correct
reconstruction, and therefore determine the time delay, by checking the
consistency of the reconstructed individual lightcurves with the
additional interferometric observations. We performed extensive
numerical simulations of synthetic lightcurves to investigate the
dependence of the performance of this method on various parameters
which are involved in the problem. Probably the most promising
candidates for applying the method (and also for determining the
Hubble constant) are lens systems consisting of multiply imaged
compact sources and an Einstein ring, such as B0218+357 from which some
of the parameters used for our simulations were adopted.

%
% chapter1     Introduction
%
\section{Introduction}
Already many years ago it was realised (Refsdal 1964) that 
gravitational lensing provides a way to determine the Hubble
constant. This method is based on measuring the time delay $\Delta t$
between the arrival times of light rays corresponding to different images in
gravitational lens systems. Of course, an intrinsic variation of the
source on appropriate timescales is essential for the time delay to be
detectable. In addition, the redshifts of source and deflector and a
reliable model for the mass distribution of the deflector are required
to calculate the Hubble constant. The advantage of the time delay
method (as also of the Sunyaev-Zel'dovich effect and potentially also of SN Ia)
compared to conventional techniques for determining $H_0$ is the
possibility to reach cosmological distances directly without using a
``distance ladder'' of intermediate calibration steps. 

The first lens system discovered, the double quasar 0957+561 (Walsh et
al. 1979), has been monitored since then at optical and radio
frequencies, and quite a number of papers about its time delay have
been published. However, the value of $\Delta t$ still remains
controversial today, because these studies were plagued by problems
like unevenly sampled data points and microlensing effects. For an
overview of the {\it status quo}\/ see Chap.~2 of Kochanek \& Hewitt (1996).  
But even if a secure value for $\Delta t$ were available, this would
not be of much use for determining the Hubble constant, because it is
extremely difficult to specify a unique model for the lens mass
distribution in this system. This problem arises because not only a single
galaxy, but also the cluster of galaxies in which it is located, as
well as another cluster at a different redshift, are contributing to the image
splitting, and so even in the simplest realistic models there are more
parameters than can be constrained observationally (e.g. Kochanek
1991, Bernstein et al. 1993; but also see Grogin \& Narayan 1996).   

A very much better candidate for determining $H_0$ is the lens system
B0218+357 (Patnaik et al. 1993), consisting of two images of a
compact flat spectrum radio source and an Einstein ring of extended
emission. This system combines several important advantages. 
The small image separation of $0''\!.335$ implies that
probably a single galaxy is acting as a lens. In addition, the
morphology of the compact images (Patnaik et al. 1995) and
especially the radio ring provide valuable information about the mass
distribution of the lensing galaxy. Thus it should be feasible to
construct a sufficiently accurate and well-constrained lens model for this
system. Furthermore, as the time delay for this system is
expected to be of the order of two weeks, and the source is known to be
variable, a determination of $\Delta t$ should be possible on reasonable
timescales of a few months. In fact, Corbett et al. (1996) already
published a value of $\Delta t=12\pm 3$ days, derived from the time
variation of the image polarization in VLA monitoring observations.   

In this paper we are investigating the possibilities for determining
the time delay $\Delta t$ of radio-loud gravitational lens systems
without the excessive use of expensive interferometric
observations. Instead, we consider monitoring the total flux density of the
unresolved lens system with a single-dish radio telescope. As we
briefly motivate in Sect.~\ref{autocorrelation} an analysis of the
autocorrelation function of the combined lightcurve is a straightforward
approach to extract the value of the time delay. However, numerical
simulations show that it is difficult to apply this method in practice,
because it would require unrealistically long monitoring periods in
order to achieve significant results. Therefore the main part of this
paper (Sect.~\ref{lightcurve}) is devoted to a new method which is
shown to yield a much more reliable result for the time delay by
making use of a few additional interferometric measurements. The idea
of this method is to reconstruct the lightcurves of the individual
images from the observed combined lightcurve by assuming values for
the time delay and the magnification ratio of the images. To single
out the true value of the time delay we use the consistency of the
reconstructed lightcurves with additional information about the flux
density ratio of the images at different epochs obtained from
interferometric observations. In extensive simulations we investigate
the performance of this ``lightcurve reconstruction method'',
depending on the various parameters involved, and show that 
typically a handful of interferometric observations suffice to determine
$\Delta t$ reliably. Finally, in Sect.~\ref{discussion} we summarize
the results and discuss the prospects for the application of this new
method, especially in context with the most promising lens system B0218+357. 

The discussion in this paper is restricted to gravitational lens
systems with two images of a compact source, although the formalism
described here might as well be extended to a larger number of
images. As we are mainly interested in radio lightcurves it is
justified to neglect microlensing effects for this study. Furthermore
we assume that there is no variation on relevant timescales in the
flux density of extended emission which may be associated with the compact
source (e.g. the ring in B0218+357). 

\newpage

%
% chapter2
%
\section{Autocorrelation approach}\label{autocorrelation}
\subsection{Time delay effect on the autocorrelation function}
The total lightcurve $\tilde S(t)$ of a two-image gravitational lens
system results from a time shifted superposition of the intrinsic
source lightcurve $\tilde S_\indrm {in}(t)$ and a constant contribution
$\tilde S_\indrm {const}$ (e.g., from an extended source component) in
the following way: 
\begin{equation}\label{superposition1}
\tilde S(t)=\tilde S_1 (t)+\tilde S_2 (t)+\tilde S_\indrm {const}=
\mu_1\tilde S_\indrm {in}(t)+\mu_2\tilde S_\indrm {in}(t-\Delta t)+
\tilde S_\indrm {const}~.
\end{equation}
$\tilde S_{1,2} (t)$ and $\mu_{1,2}$ denote the flux density and the
absolute magnification of the individual images.\footnote{In the following
we will restrict the time delay to be positive. Hence the index $1$ denotes
the image in which intrinsic source variations will appear
first.} Introducing the observable magnification ratio
$\mu:={\mu_1}/{\mu_2}$ and subtracting the time averages of the
flux densities, $S(t):=\tilde S(t)-\langle\tilde S(t)\rangle_t$,
$S_1(t):=\tilde S_1(t)-\langle\tilde S_1(t)\rangle_t$ and 
$S_2(t):=\tilde S_2(t)-\langle\tilde S_2(t)\rangle_t$,
Eq.~(\ref{superposition1}) becomes  
\begin{equation}\label{superposition2}
S(t)=S_1(t)+S_2(t)=S_1(t)+\frac{1}{\mu}S_1(t-\Delta t)~.
\end{equation}
Here we assume that $S(t)$ is a stationary random process and define its
(normalized) autocorrelation function $C(\tau)$ as
\[
C(\tau):=\frac{\xi(\tau)}{\xi(0)}\ \ \ \mbox{with}\ \ \ 
\xi(\tau):=\,\langle S(t)\,S(t+\tau)\rangle_t~.
\]
It is obvious that the time delay should show up in $C(\tau)$ as a
positive contribution to the autocorrelation at $\tau=\Delta t$.
Starting from Eq.~(\ref{superposition2}) it is an easy exercise to show
that $C(\tau)$ can be calculated from the intrinsic autocorrelation function
$C_\indrm {in}(\tau)$ of the source according to
\begin{equation}\label{automodif}
C(\tau)=\frac{1}{\alpha}\left[C_\indrm {in}(\tau)
+\frac{\mu}{\mu^2+1}C_\indrm {in}(\tau-\Delta t)
+\frac{\mu}{\mu^2+1}C_\indrm {in}(\tau+\Delta t)\right]~,
\end{equation}
with the factor $\alpha=1+\frac{2\mu}{\mu^2+1}C_\indrm {in}(\Delta t)$
ensuring the proper normalization. Corresponding to the change in the
autocorrelation function the power spectrum is modified according to 
\begin{equation}\label{powermodif}
P(\omega)=\mu_1^2\,P_\indrm{in}(\omega)
\left[\left(1+\frac{1}{\mu^2}\right)+\frac{2}{\mu}\cos(\omega\Delta t)\right]~,
\end{equation}
where $P_\indrm{in}(\omega)$ denotes the intrinsic power spectrum of the
unlensed source.

\begin{figure}[t]
  \vspace{-3mm}
    \leavevmode
    \epsfxsize=8cm
    \epsffile{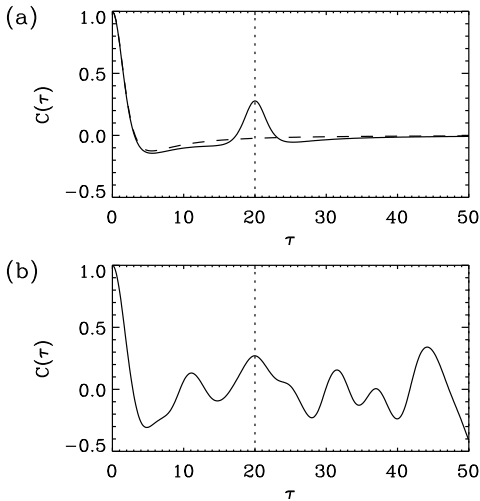}
  \vspace{-6mm}
\parbox[b]{7.8cm}{
  \renewcommand{\baselinestretch}{0.8}
  \caption{\footnotesize {\bf(a)} The autocorrelation function $C(\tau)$
(solid line) resulting from a time shifted superposition of an
intrinsic lightcurve with autocorrelation function
$C_\indfrm{in}(\tau)$ (dashed line). The value of $\Delta t$ is
indicated by the dotted line. {\bf(b)} This diagram shows the
autocorrelation function calculated for a simulated lightcurve with
parameters as shown in Table~\protect\ref{lightstandard}. Again the
dotted line indicates the value for the time delay.}   
  \label{autocorr}
  \vspace{3mm}   }
\end{figure}

Figure~\ref{autocorr}a graphically depicts the modification of the
autocorrelation function described by Eq.~(\ref{automodif}). Here we took the
values $\Delta t=20$ (in arbitrary time units\footnote{By scaling all
time variables appropriately the arbitrary time units used here can be
adapted to any lens system of interest.}) for the time delay, $\mu=3$
for the magnification ratio and 
\begin{equation}\label{autostandard}
C_\indrm{in}(\tau)=\tau_\indrm{var}^2\frac{\tau_\indrm{var}^2-\tau^2}
{\left(\tau_\indrm{var}^2+\tau^2\right)^2}~
\end{equation}
as an example for the intrinsic autocorrelation function, which
corresponds to the reasonable intrinsic power spectrum
\begin{equation}\label{powerspectrum}
P_\indrm{in}(\omega)
\propto\omega\,{\mbox e}^{-\textstyle\tau_\indfrm{var}\,\omega}~. 
\end{equation}
The parameter $\tau_\indrm{var}$ represents a characteristic timescale
of the intrinsic source variability and has been chosen to be
$\tau_\indrm{var}=3.18$ for this plot. The resulting autocorrelation
function $C(\tau)$ shows a distinct maximum at the value of the time
delay. Thus it is at least in principle possible to obtain information
about $\Delta t$ by analysing the autocorrelation function of combined
lightcurves. However, if the timescale of variability
$\tau_\indrm{var}$ is comparable to or longer than $\Delta t$, the
``time delay peak'' will merge with the ``intrinsic maximum'' of
$C(\tau)$ at $\tau=0$ and any information on $\Delta t$ will be lost. 
\subsection{Analysis of $C(\tau)$ for simulated lightcurves}
Of course, in applications to observations the limited observing time $T$,
the finite sampling interval $\Delta T$ and observational errors will
impose serious constraints on the usefulness of the autocorrelation
function for determining the time delay. To study this quantitatively
we used synthetic data sets of combined lightcurves. These lightcurves
were generated as realizations of a gaussian random process with
the intrinsic power spectrum of Eq.~(\ref{powerspectrum}). For
simplicity we restricted the study to constant observing intervals $\Delta T$. 
\begin{table}[b]
  \vspace{-3mm}
  \renewcommand{\baselinestretch}{0.8}
  \caption{\footnotesize Standard parameter values for simulated
lightcurves. The parameter $\eta$ is defined as the ratio of the
standard deviation $\sigma_{\delta S}$ of observational errors added
to every data point and the dispersion $\sigma_S$ of the lightcurve itself.}  
  \label{lightstandard}
  \vspace{-1mm}
  \begin{center}
\begin{tabular}{|l|c|c|} \hline
parameter&symbol&standard value \\ \hline\hline
observing period&$T$&100 \\ \hline
observing interval&$\Delta T$&1 \\ \hline
observing error&$\eta$&0.1 \\ \hline
power spectrum&$P_\indrm{in}(\omega)$&
$\propto\omega\,{\mbox e}^{-\textstyle\tau_\indfrm {var}\,\omega}$ \\ \hline
timescale of variability&$\tau_\indrm{var}$&3.18 \\ \hline
magnification ratio&$\mu$&3 \\ \hline
time delay&$\Delta t$&20 \\ \hline
\end{tabular}
  \end{center}
  \vspace{-4mm}
\end{table}

In Fig.~\ref{autocorr}b the autocorrelation function for one typical
lightcurve realization with parameters as summarized in
Table~\ref{lightstandard} is plotted. Due to the limited observing 
period $T$, which does not provide a ``fair sample'' of the
lightcurve's statistical properties, a number of additional maxima and
minima are now showing up, making it very difficult to identify the
time delay peak. A simple quantitative measure for the performance of
the autocorrelation method is the fraction $\cal{P}$ of lightcurve
realizations for which the time delay maximum, i.e. the maximum
closest to the true time delay, is in fact the highest maximum within
the range of potential $\Delta t$ values. In the following all maxima
in the range $[0,50]$ except the $\tau$=0~maximum are considered. We
studied the dependence of $\cal{P}$ (and other measures) on all of the
parameters shown in Table~\ref{lightstandard}, but here we only
present the results for varying $T$ and $\mu$, keeping the other
parameters fixed. 
\begin{figure}[t]
  \vspace{-3mm}
    \leavevmode
    \epsfxsize=8cm
    \epsffile{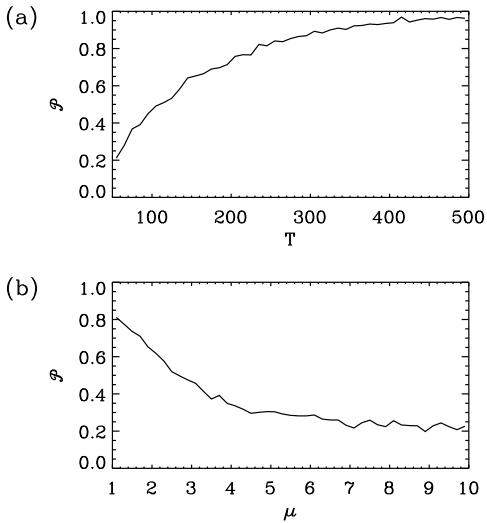}
  \vspace{-0mm}
\parbox[b]{7.8cm}{
  \renewcommand{\baselinestretch}{0.8}
  \caption{\footnotesize The dependence of $\cal{P}$ (see text for
definition) on {\bf(a)} the observing period $T$ and {\bf(b)} the
magnification ratio $\mu$ of the images; all other parameters are as
in Table~\protect\ref{lightstandard}.}  
  \label{autoTmue}
  \vspace{3mm}   }
\end{figure}

From Fig.~\ref{autoTmue}a it can be seen that prolonging the
observation period $T$ leads to an increase in $\cal{P}$, because with
increasing $T$ the sampling of the lightcurve improves and the
observed autocorrelation function converges to the theoretical
$C(\tau)$ shown in Fig.~\ref{autocorr}a. However, since for some
realizations the unwanted additional maxima in $C(\tau)$ can be quite stable,
it takes a rather long observing time to ensure getting significant
results for the time delay. In Fig.~\ref{autoTmue}b the dependence of
$\cal{P}$ on the magnification ratio is depicted. Of course the method
works best for $\mu=1$, as for increasing magnification ratio the
combined lightcurve is dominated more and more by the brighter image.
The autocorrelation function $C(\tau)$ will then be dominated by the
first term in Eq.~(\ref{automodif}), which does not contain the time
delay, and converges to the intrinsic autocorrelation function
$C_\indrm{in}(\tau)$ for $\mu\rightarrow\infty$.  

To summarize, one can say that an analysis of the autocorrelation
function of the total lightcurve of gravitational lens systems only
constitutes a viable method for determining the time delay if the
magnification ratio is close to unity and the observing period $T$ is
{\it very much} longer than $\Delta t$ and $\tau_\indrm{var}$.

%
% chapter3
%
\section{Lightcurve reconstruction method}\label{lightcurve}
In this section we introduce the ``lightcurve reconstruction method''
for measuring the time delay of gravitational lens systems. As we will 
explain in Sect.~\ref{reconstruction}, it is possible to reconstruct 
the (unobserved) lightcurves $S_1(t)$ and $S_2(t)$ of the individual
images, if we assume values for the time delay $\Delta t$ and the
magnification ratio $\mu$. Sect.~\ref{constraints} shows qualitatively
how the true $\Delta t$ value can then be singled out by checking the
consistency of the reconstructed lightcurves with additional
information from interferometric measurements about the flux density
ratio of the images. In Sect.~\ref{chisqu} this is discussed
quantitatively in terms of the $\chi^2$ function. Using lightcurve
simulations we investigate in Sect.~\ref{parameters} the dependence of
the results on the various parameters that are involved in this
problem. Finally, Sect.~\ref{minim} shows how confidence intervals for
the time delay can be obtained in individual realizations.   
\subsection{Reconstruction of lightcurves}\label{reconstruction}
\begin{figure}[t]
  \vspace{-3mm}
  \begin{center}
    \leavevmode
    \epsfxsize=15.9cm
    \epsffile{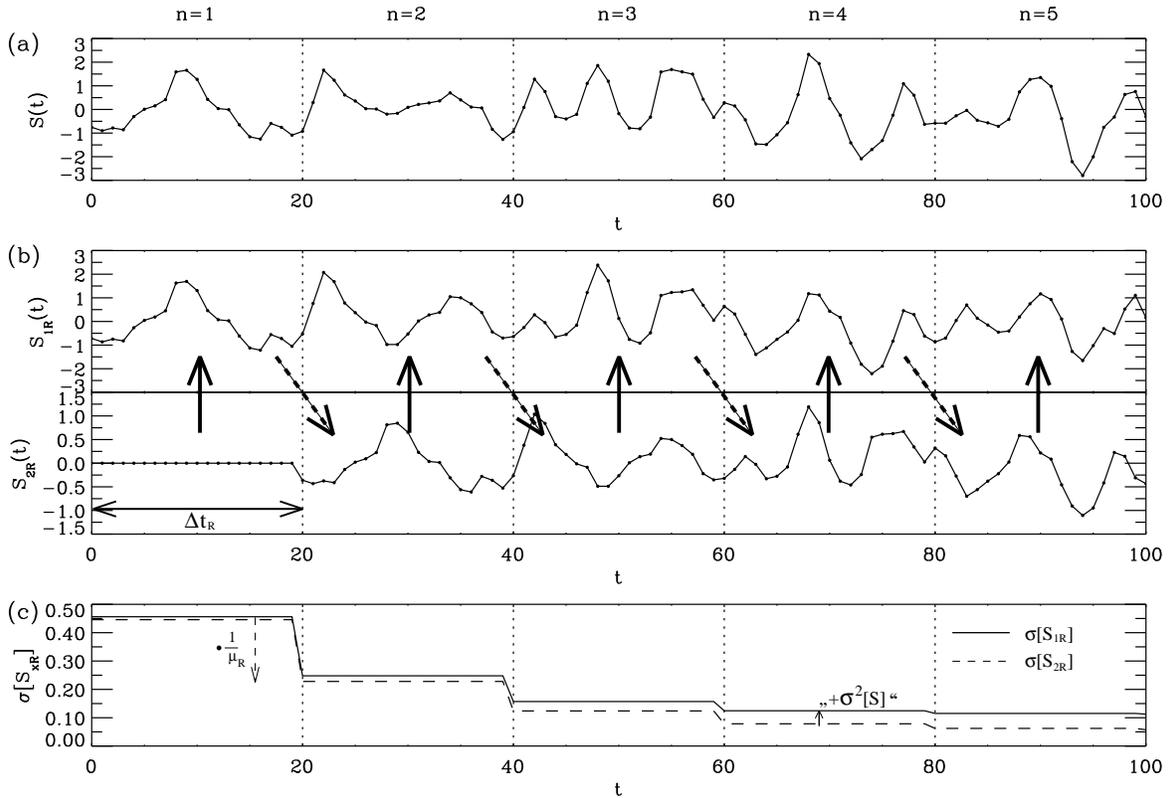}
  \end{center}
  \vspace{-2mm}
  \renewcommand{\baselinestretch}{0.8}
  \caption{\footnotesize The principle of lightcurve reconstruction:
{\bf(a)} The observed combined lightcurve. {\bf(b)} Reconstruction of
the individual lightcurves with $\dtrf=20$ and $\murf=2$ by iteratively
applying Eq.~(\protect\ref{subtract}) (solid arrows) and
Eq.~(\protect\ref{shift}) (dashed arrows). {\bf(c)} The errors
$\sig{S_\indrm{1R}}$ and $\sig{S_\indrm{2R}}$ of the reconstructed
individual lightcurves under the hypothesis that $\dtrf$ and $\murf$ are
in fact the true values. These were calculated from
Eqs.~(\protect\ref{erriter2}) and (\protect\ref{erriter1}) with
constant observing error $\sig{S^{(n)}}=\sigma_{\delta S}=\eta\,\sigma_S$ 
and using an estimate for the initial errors as described in the text.}  
  \label{ReconstDemo}
\end{figure}

It is convenient to discuss the method by looking at Fig.~\ref{ReconstDemo}.  
As an example for an ``observed'' combined lightcurve
Fig.~\ref{ReconstDemo}a displays one realization of a gaussian random
process with parameters as shown in Table~\ref{lightstandard} except
that here we took $\mu=2$. Now, to reconstruct the individual image
lightcurves $S_\indrm{1R}(t)$ and $S_\indrm{2R}(t)$ we have to
postulate values $\dtr$ and $\mur$ for the time delay and the
magnification ratio, respectively.\footnote{Here and in the following the
subscript $_\indfrm{R}$ indicates quantities used for or obtained from the
reconstruction.} It will become clear in the following subsections how
to actually determine these quantities with this method. In the
example of Fig.~\ref{ReconstDemo} we used just for illustration the
values $\dtr=20$ and $\mur=2$ which in fact agree with the ``true'' values
used for generating the combined lightcurve. In addition, we have to
make a guess about the individual lightcurve $S_\indrm{2R}(t)$ in the time
interval $t\in[0,\dtr[$, but luckily the choice for this initial guess
will turn out to be rather irrelevant for the method to work. For
simplicity we use $S_\indrm{2R}(t)=0$ for $t\in[0,\dtr[$ within this paper.   

Having specified $\dtr$, $\mur$ and $S_\indrm{2R}(t)$ for $t\in[0,\dtr[$ we can
immediately calculate the individual lightcurve $S_\indrm{1R}(t)$ in the same
interval using
\begin{equation}\label{subtract}
S_\indrm{1R}(t)=S(t)-S_\indrm{2R}(t)
\end{equation}
which is trivially derived from Eq.~(\ref{superposition2}). The application of
Eq.~(\ref{subtract}) is visualized by a solid arrow in Fig.~\ref{ReconstDemo}b.
The next step is to compute $S_\indrm{2R}(t)$ on the subsequent
interval $[\dtr,2\dtr[$ using
\begin{equation}\label{shift}
S_\indrm{2R}(t)=\frac{1}{\mur}S_\indrm{1R}(t-\dtr)~.
\end{equation}
This is indicated by the dashed arrows of Fig.~\ref{ReconstDemo}b. By
iteratively applying Eqs.~(\ref{subtract}) and (\ref{shift}), the individual
image lightcurves can be reconstructed for the total observing period.
Introducing the notation $S^{(n)}$ for the restriction of the function
$S(t)$ to the interval $[(n-1)\,\dtr,n\,\dtr[$, the reconstruction process
can be written as\footnote{Note that the quantities in this notation
are still meant to be functions of $t$ although the argument has been
omitted.} 
\begin{equation}\label{iteration2}
S^{(n)}_\indrm{2R}=\sum_{m=1}^{n-1}\frac{(-1)^{m+1}}{\mur^m}S^{(n-m)}
+\frac{(-1)^{n-1}}{\mur^{n-1}}S^{(1)}_\indrm{2R}
\end{equation}
and
\begin{equation}\label{iteration1}
S^{(n)}_\indrm{1R}=\sum_{m=0}^{n-1}\frac{(-1)^m}{\mur^m}S^{(n-m)}
+\frac{(-1)^n}{\mur^{n-1}}S^{(1)}_\indrm{2R}~.
\end{equation}

Now we would like to quantify the rms errors $\sig{S^{(n)}_\indrm{1R}}$ 
and $\sig{S^{(n)}_\indrm{2R}}$ of the reconstructed individual
lightcurves under the hypothesis that the values $\dtr$ and $\mur$
used for the reconstruction are in fact the true ones. Therefore these
errors include the observational errors of the total lightcurve
measurements and the errors caused by the arbitrary initial guess. In
the following we assume the observational errors to be normally
distributed with standard deviation $\sig{S^{(n)}}$ and statistically
independent for each data point. Hence the increase of
$\sig{S^{(n)}_\indrm{1R}}$ with respect to $\sig{S^{(n)}_\indrm{2R}}$ 
when applying Eq.~(\ref{subtract}) is described by   
\begin{equation}\label{errsubtract}
\sigs{S^{(n)}_\indrm{1R}}=\sigs{S^{(n)}_\indrm{2R}}+\sigs{S^{(n)}}~,
\end{equation}
whereas using Eq.~(\ref{shift}) leads to an evolution of the error
according to
\begin{equation}\label{errshift}
\sig{S^{(n+1)}_\indrm{2R}}
=\frac{1}{\mur}\,\sig{S^{(n)}_\indrm{1R}}~.
\end{equation}
In analogy to Eqs.~(\ref{iteration2}) and (\ref{iteration1}) the
error propagation during the reconstruction can be compactly expressed as
\begin{equation}\label{erriter2}
\sigs{S^{(n)}_\indrm{2R}}=\sum_{m=1}^{n-1}\frac{1}{\mur^{2m}}
\sigs{S^{(n-m)}}+\frac{1}{\mur^{2(n-1)}}\sigs{S_\indrm{2R}^{(1)}}~
\end{equation}
and
\begin{equation}\label{erriter1}
\sigs{S^{(n)}_\indrm{1R}}=\sum_{m=0}^{n-1}\frac{1}{\mur^{2m}}
\sigs{S^{(n-m)}}+\frac{1}{\mur^{2(n-1)}}\sigs{S_\indrm{2R}^{(1)}}~.
\end{equation}
However, looking at Fig.~\ref{ReconstDemo}c is much more illustrative
to understand the evolution of the errors of the reconstructed
lightcurves. For the first interval $[0,\dtr[$ we have to estimate the error
$\sig{S_\indrm{2R}^{(1)}}$ caused by the arbitrary initial guess. Here
we assume that the variability characteristics of the second lightcurve in the
starting interval does not substantially differ from the observed
variability of the combined lightcurve during the total observing
period. This leads to the estimate
$\sigs{S^{(1)}_\indrm{2R}}\approx\frac{1}{1+\murf^2}\sigma^2_{S}$ with
$\sigma_{S}$ denoting the dispersion of the combined lightcurve during
the observation period.\footnote{For lightcurves with timescale of
variability comparable to (or larger than) $\dtrf$ the dispersion is
increasing when extending the interval on which it is calculated from
$\dtrf$ to $T$. Therefore, the dispersion used for this estimate should be
calculated on an interval equal to $\dtrf$ in such cases, because
otherwise the error of the initial guess would be overestimated.}     
Figure~\ref{ReconstDemo}c clearly shows that the initial uncertainty in the
reconstructed lightcurves decreases at every interval, because the errors are
multiplied with the factor $1/\murf$ in Eq.~(\ref{errshift}). Thus the
reconstruction is most accurate at the end of the observing
period. Here the errors are dominated by the observational errors of the
combined lightcurve, which are added when applying
Eq.~(\ref{errsubtract}), and the initial uncertainty has almost dropped out.
Of course the reconstruction works best for large values of the
magnification ratio $\mur$ because then the initial errors are 
decreasing faster. In the following we will restrict $\mur$ to be
larger than one, because otherwise the errors will actually grow and
the reconstruction fails. However, this is no fundamental problem for the
method, since for $\mur<1$, i.e. in cases in which the light arrives
first in the weaker image, the reconstruction can be done starting at
the end of the observing period.  

For simplicity we will assume for this study that the standard
deviation of the observational errors of the combined lightcurve is
constant for all data points, $\sig{S^{(n)}}=\sigma_{\delta S}$. In
addition, we restrict the simulations performed in this paper to constant 
observing intervals $\Delta T$. It should be emphasized here that the
limitation to constant $\sigma_{\delta S}$ and $\Delta T$ is merely
for convenience, and removing these restrictions does not impose any conceptual
difficulties for the reconstruction method. An important point to be
mentioned is the following. The errors $\sig{S_\indrm{1R}}$ and
$\sig{S_\indrm{2R}}$ shown in Fig.~\ref{ReconstDemo}c are strictly
correct only for integer multiples of the observing interval $\Delta T$. In
order to calculate the reconstructed lightcurves for $t$ values lying
between the data points, the observed combined lightcurve has to be
interpolated which in general introduces additional errors. 
Interpolation also is necessary to do the reconstruction with $\dtr$
not being an integer multiple of $\Delta T$. We use a linear
interpolation and determine the error introduced by the interpolation
directly from an ensemble of simulated lightcurves. For real data,
however, an extrapolation of the variability characteristics to
timescales shorter then the sampling interval $\Delta T$ has to be
made in order to get an estimate of the errors introduced by
interpolation. In practice this could be achieved by following the
approach of Press et al. (1992a). They describe a method that provides
a ``reconstruction of a set of irregularly sampled measurements into a
continuous function and an associated standard error function'' by
using an estimate of the underlying autocorrelation function obtained
from the data.  

\begin{figure}[t]
  \vspace{-0mm}
    \leavevmode
    \epsfxsize=10cm
    \epsffile{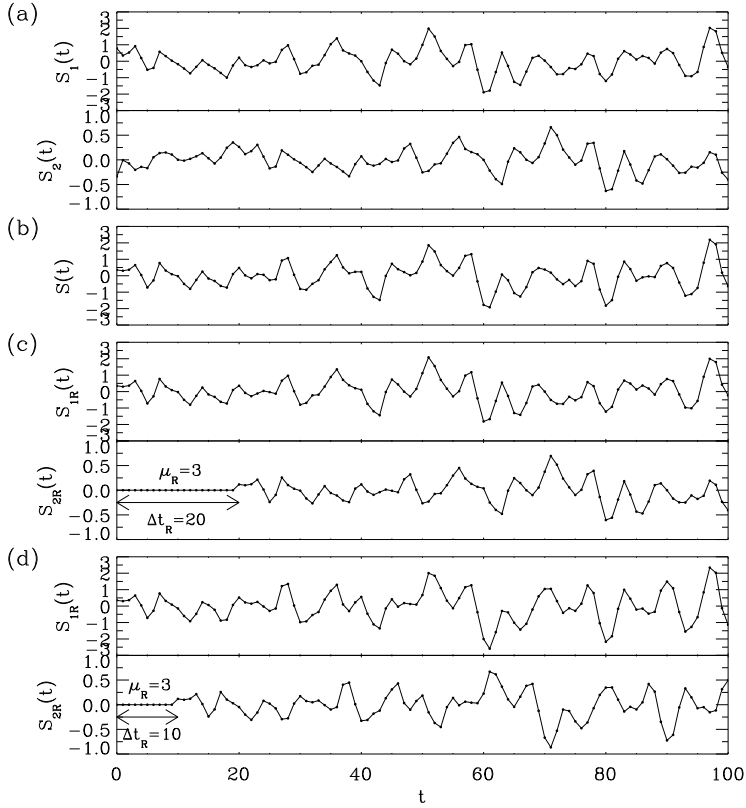}
  \vspace{2mm}
\parbox[b]{5.8cm}{
  \renewcommand{\baselinestretch}{0.8}
  \caption{\footnotesize One realization of a gaussian random process
generated with parameters according to Table~\protect\ref{lightstandard}. 
{\bf(a)} The {\it unobserved}\/ individual lightcurves $S_1(t)$ and
$S_2(t)$. {\bf(b)} The {\it observed}\/ combined lightcurve that results
by adding up $S_1(t)$ and $S_2(t)$ with $\Delta t=20$ and $\mu=3$.
{\bf(c)} The {\it reconstructed}\/ individual lightcurves
$S_\indfrm{1R}(t)$ and $S_\indfrm{2R}(t)$ using the {\it true}\/
values $\dtrf=20$ and $\murf=3$. {\bf(d)} The reconstruction when using 
$\murf=3$ and the {\it wrong}\/ value $\dtrf=10$ for the time delay.} 
  \label{RichtigFalsch}
  \vspace{2mm}   }
\end{figure}

Figure~\ref{RichtigFalsch} shows another example for a lightcurve
generated according to a gaussian random process\footnote{Note that
the reconstruction method does not rely on properties like gaussianity
or stationarity. The reason to use gaussian random processes for the
simulations is just because they are easy to generate.} with the parameters of
Table~\ref{lightstandard}. In Fig.~\ref{RichtigFalsch}a the unobserved
individual image lightcurves $S_1(t)$ and $S_2(t)$ are depicted. With
$\Delta t=20$ and $\mu=3$ these add up to the observed combined
lightcurve $S(t)$ in Fig.~\ref{RichtigFalsch}b. In Fig.~\ref{RichtigFalsch}c 
the individual lightcurves are reconstructed using the true values for
$\dtr$ and $\mur$, whereas Fig.~\ref{RichtigFalsch}d shows the
reconstruction using the wrong value $\dtr=10$ for the time delay.
In agreement with the error discussion above the correctly
reconstructed lightcurves resemble the true ones quite closely towards
the end of the observing period. Of course now the problem arises how
to distinguish between the correct reconstruction
(Fig.~\ref{RichtigFalsch}c) and the wrong one in Fig.~\ref{RichtigFalsch}d 
as well as all the other wrong reconstructions conceivable with
different $\dtr$'s and $\mur$'s. Although it might be possible in some
special cases to infer $\Delta t$ by making assumptions about the
shape of correctly reconstructed lightcurves, this is certainly not
viable in general. Therefore we clearly need some additional
information to solve the problem.
\subsection{Flux density ratios as additional constraints}\label{constraints}
For the rest of this paper we investigate the chances for determining
the time delay by using additional observational constraints about the
flux density ratio of the images obtained from interferometric observations.
It is useful to take the flux density ratio rather than absolute flux
densities of the images, because measuring the former does not suffer from
calibration problems and is therefore much more accurate.

The time dependent flux density ratio $m(t)$ of the two images is given by
\begin{equation}\label{flux}
m(t):=\frac{\tilde S_1(t)}{\tilde S_2(t)}=\frac
{\langle\tilde S_1(t)\rangle_t+S_1(t)}{\langle\tilde S_2(t)\rangle_t+S_2(t)}
=\frac{\langle\tilde S_{1+2}\rangle\,\frac{\mu}{1+\mu}
+S_1(t)}{\langle\tilde S_{1+2}\rangle\,\frac{1}{1+\mu}+S_2(t)}~,
\end{equation}
where $\tilde S_1(t)$ and $\tilde S_2(t)$ denote the flux densities
without subtraction of the mean values (cf. the definition in
Sect.~\ref{autocorrelation}). In the final identity of this equation
the individual image time averages have been expressed in terms of the
combined lightcurve average 
$\langle\tilde S_{1+2}\rangle:=\langle\tilde S_1(t)+\tilde S_2(t)\rangle_t$. 
Note that $\langle\tilde S_{1+2}\rangle$ only includes the flux densities of
the varying components in the lens system. Thus in order to determine
$\langle\tilde S_{1+2}\rangle$ at least one interferometric observation is
required to subtract the constant flux density of non varying
components from the average single-dish flux density.

In analogy to Eq.~(\ref{flux}) we can calculate the reconstructed flux
density ratio $m_\indrm{R}(t)$ from the reconstructed individual image
lightcurves as follows:
\begin{equation}\label{recflux}
m_\indrm{R}(t)=
\frac{\langle\tilde S_{1+2}\rangle\,\frac{\murf}{1+\murf}
+S_\indrm{1R}(t)}{\langle\tilde S_{1+2}\rangle\,
\frac{1}{1+\murf}+S_\indrm{2R}(t)}= 
\frac{\langle\tilde S_{1+2}\rangle\,\frac{\murf}{1+\murf}
+S(t)-S_\indrm{2R}(t)}{\langle\tilde S_{1+2}\rangle\,
\frac{1}{1+\murf}+S_\indrm{2R}(t)}~. 
\end{equation}
In order to express $m_\indrm{R}(t)$ in terms of quantities with
independent errors we replaced $S_\indrm{1R}(t)$ with
$S(t)-S_\indrm{2R}(t)$ in this equation, because for a given epoch $t_i$ the
reconstruction error $\sig{S_\indrm{2R}(t_i)}$ only contains
observational errors of $S(t)$ for epochs preceding $t_i$ and
therefore is independent of $\sig{S(t_i)}$. By applying the usual
error propagation law we obtain the relation
\begin{eqnarray}\label{flussfehler}
\sigs{m_\indrm{R}(t)}&=&
\frac{\left[\frac{\murf}{1+\murf}S_\indrm{2R}(t)-
\frac{1}{1+\murf}S_\indrm{1R}(t)\right]^2}{\left[\langle\tilde S_{1+2}\rangle
\frac{1}
{1+\murf}+S_\indrm{2R}(t)\right]^4}\,\sigs{\langle\tilde S_{1+2}\rangle}+
\frac{1}
{\left[\langle\tilde S_{1+2}\rangle\frac{1}{1+\murf}+S_\indrm{2R}(t)\right]^2
}\,\sigs{S(t)}+\nonumber\\&~&
+\frac{\left[\langle\tilde S_{1+2}\rangle+S_\indrm{1R}(t)+
S_\indrm{2R}(t)\right]^2}{\left[\langle\tilde S_{1+2}\rangle
\frac{1}{1+\murf}+S_\indrm{2R}(t)\right]^4}\,\sigs{S_\indrm{2R}(t)}
\end{eqnarray}
for the error $\sig{m_\indrm{R}(t)}$ of the reconstructed flux density
ratio under the hypothesis that $\dtr$ and $\mur$ are the
true values.
\begin{figure}[b]
  \vspace{+2mm}
    \leavevmode
    \epsfxsize=10cm
    \epsffile{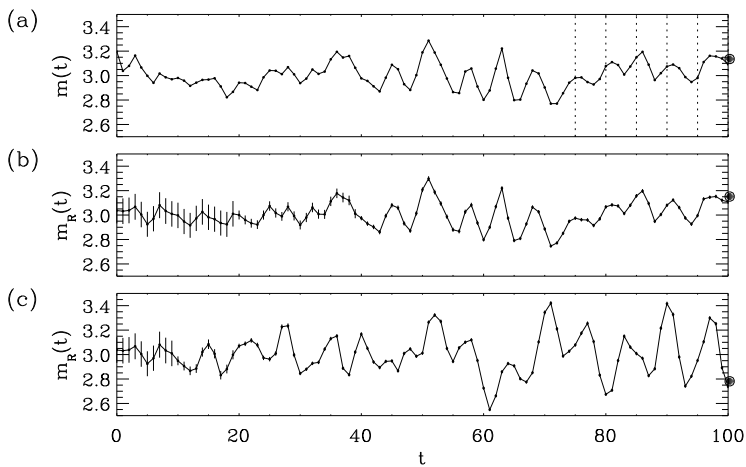}
  \vspace{+2mm}
\parbox[b]{5.8cm}{
  \renewcommand{\baselinestretch}{0.8}
  \caption{\footnotesize {\bf(a)} The {\it unobserved}\/ time
evolution of the flux density ratio $m(t)$ for the example lightcurve plotted
in Fig.~\protect\ref{RichtigFalsch}. {\bf(b)} The flux density ratio
$m_\indfrm{R}(t)$ reconstructed with the {\it true}\/ values $\dtrf=20$ and
$\murf=3$. {\bf(c)} The {\it wrong}\/ reconstruction with $\dtrf=10$
and $\murf=3$.
\protect\newline
The filled circles at $t=100$ and the dotted vertical
lines in the top panel indicate additional interferometric flux
density ratio measurements.}  
  \label{FlussZeit}
  \vspace{2mm}   }
\end{figure}

Figure~\ref{FlussZeit} shows the unobserved true time evolution of the
flux density ratio $m(t)$ for the example of Fig.~\ref{RichtigFalsch}
as well as its correct and wrong reconstructions $m_\indrm{R}(t)$,
corresponding to the lightcurve reconstructions in Fig.~\ref{RichtigFalsch}c
and d. In order to compute the flux density ratio curves for this example we
have to specify the value of $\langle\tilde S_{1+2}\rangle$ appearing
in Eqs.~(\ref{flux}) and (\ref{recflux}). To do so we introduce a new
parameter for our lightcurve simulations, which is not included in
Table~\ref{lightstandard} because it was not relevant for the
autocorrelation analysis. We define the parameter
\[
\nu:=\frac{\sigma_S}{\langle\tilde S_{1+2}\rangle}
\]
as the ratio of the total lightcurve's dispersion to the average
combined flux density in order to describe the relative variability of the
source. For the example case a variability of $\nu=2\%$ has been chosen.
The reconstructed flux density ratio curves in Fig.~\ref{FlussZeit}
also include $1\sigma$ error bars calculated from
Eq.~(\ref{flussfehler}). Again it can be seen that the initial
uncertainty gradually decreases during the reconstruction process.
At the end of the observing period ($t=100$) the contributions of
$\sig{S(t)}$ and $\sig{S_\indrm{2R}}$ to the error of the flux density
ratio are comparable, whereas that of $\sig{\langle\tilde S_{1+2}\rangle}$ is
negligible.
\begin{figure}[t]
  \vspace{-2mm}
    \leavevmode
    \epsfxsize=15.9cm
    \epsffile{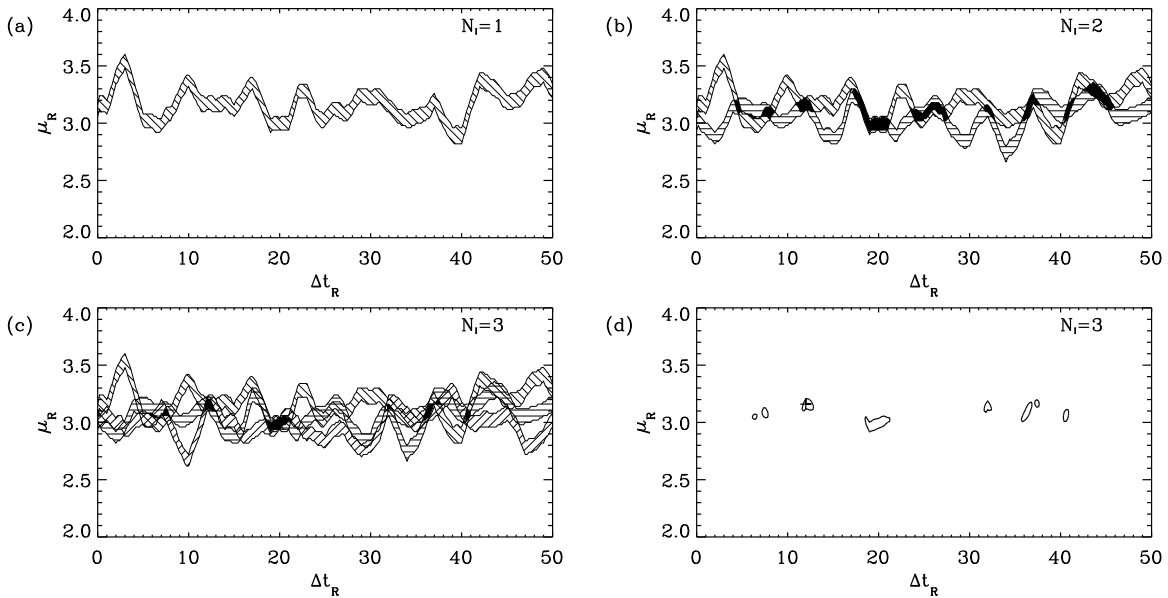}
  \vspace{-1mm}
  \renewcommand{\baselinestretch}{0.8}
  \caption{\footnotesize {\bf(a)} The hatched area includes all pairs
of values for $\dtrf$ and $\murf$ with the reconstructed flux density
ratio $m_\indrm{R}(t_1=100)$ being compatible with the actually
observed value $m(t_1=100)$ within a $3\sigma$ error interval. In
{\bf(b)} and {\bf(c)} the regions of $3\sigma$ consistency with flux
density ratio constraints at $t_2=95$ and $t_3=90$ have been
added. $N_\indfrm{I}$ denotes the number of interferometric
measurements. Corresponding to the diagram in {\bf(c)}, {\bf(d)} shows the
$3\sigma$ contours of the $\chi^2$ function.}  
  \label{Zusatz123}
\end{figure}

From Fig.~\ref{FlussZeit} it is clear that an additional measurement
of the flux density ratio, say at $t_1=100$, can rule out the
combination of the values $\dtr=10$ for the time delay and $\mur=3$ for
the magnification ratio, because these would lead to a completely
wrong prediction for the flux density ratio at that epoch. However, we
have to test the compatibility of all potential values for $\dtr$ and
$\mur$ with the additional constraint $m(t_1=100)$. This has been done
in Fig.~\ref{Zusatz123}a. The hatched area in this plot includes all
pairs of parameter values ($\dtr$, $\mur$) with the reconstructed flux
density ratio $m_\indrm{R}(t_1=100)$ being compatible with the
measured flux density ratio $m(t_1=100)$ within a $3\sigma$ error
margin, i.e.
\[
|m_\indrm{R}(t_1)-m(t_1)|<3\,\sqrt{\sigs{m_\indrm{R}(t_1)}+\sigs{m(t_1)}}~. 
\]
The values for the flux density ratio measurements $m(t_i)$ used here
and below include a normally distributed observational error with
standard deviation $\sig{m(t_i)}=0.005\,m(t_i)$. Note that for any time delay
$\dtr$ there is, at least for this example, a value for the
magnification ratio $\mur$ which yields a reconstruction in agreement with
the additional flux density ratio constraint. Thus in order to
restrict the range of possible time delay values, more than one flux
density ratio measurement is required. Figure~\ref{Zusatz123}b
additionally shows the $3\sigma$ consistency region defined by another
flux density ratio measurement performed at $t_2=95$, and in
Fig.~\ref{Zusatz123}c a third measurement at $t_3=90$ is included. The
intersection of these regions is filled black. These plots nicely
illustrate how the method works. However, to quantify the consistency
of parameter values ($\dtr$, $\mur$) with the interferometric
constraints a discussion in terms of $\chi^2$ is adequate.    
\subsection{Analysis of minima in the $\chi^2$ function}\label{chisqu}
In general $\chi^2$ functions are used to fit parametrized models to a
given data set and to quantify the compatibility of the model with the data.
In the case discussed here the ``model'' is represented by the basic
assumptions made for this study which were formulated in the last
paragraph of the introduction. The postulated values $\dtr$ for the
time delay and $\mur$ for the magnification ratio can be regarded as
the fit parameters. Therefore we define $\chi^2$ as a function of
$\dtr$ and $\mur$: 
\begin{equation}\label{chisquare}
\chi^2(\dtr,\mur)=\sum_{i=1}^{N_\indfrm{I}}
\frac{\left[m_\indrm{R}(t_i\,;\,\dtr,\mur)-m(t_i)\right]^2} 
{\sigs{m_\indrm{R}(t_i\,;\,\dtr,\mur)}+\sigs{m(t_i)}}~.
\end{equation}
$N_\indrm{I}$ denotes the number of interferometric flux density ratio
measurements. Note that here not only the observational data $m(t_i)$
but also the values $m_\indrm{R}(t_i\,;\,\dtr,\mur)$ predicted by the model
contain errors\footnote{These errors also depend on the fit parameters
$\dtrf$ and $\murf$. Strictly speaking, this parameter dependence of
the errors leads to an additional term which has to be considered when
deriving the $\chi^2$ function from the maximum likelihood
principle. However, this term can be neglected here because the
variation of the errors is small for parameter values of interest.}
which have to be included in the $\chi^2$ function. Another
peculiarity of the $\chi^2$ used here is the extreme non-linearity of
the model in its fit parameter $\dtr$. This leads to the ocurrence of
a variety of minima in the $\chi^2$ function and to non-elliptic
confidence regions for the parameter values. However, as we will see
below the model is still sufficiently linear locally (at individual
minima), so that $\chi^2$ statistics can be applied (see
e.g. Sect.~15.6 of Press et al. 1992b). 

\begin{figure}[b]
  \vspace{-2mm}
    \leavevmode
    \epsfxsize=8cm
    \epsffile{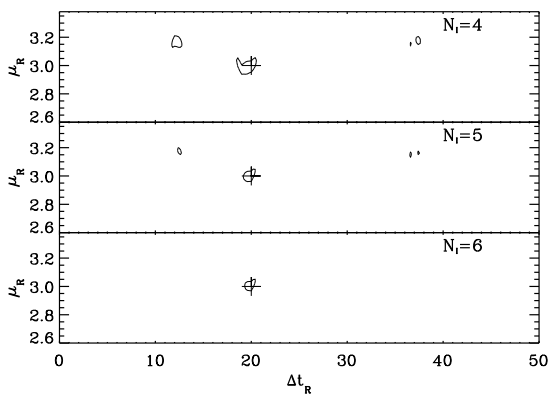}
  \vspace{2mm}
\parbox[b]{7.8cm}{
  \renewcommand{\baselinestretch}{0.8}
  \caption{\footnotesize These plots show the evolution of the
$3\sigma$ contours when increasing the number $N_\indfrm{I}$ of
interferometric observations. For $N_\indfrm{I}=6$ only the true
minimum remains. The global minimum of the $\chi^2$ function
within the parameter range considered is marked with a cross.}  
  \label{Zusatz456}
  \vspace{2mm}   }
\end{figure}

Figure~\ref{Zusatz123}d shows the $3\sigma$ contours of the
$\chi^2$ function for the same number of interferometric measurements
as in Fig.~\ref{Zusatz123}c. The definition of the $3\sigma$
consistency value for $\chi^2$ which has been used here will be given
below. This plot reveals that the information from the three flux
density ratio measurements included here does not suffice to determine
the time delay unambiguously. There are still a number of $3\sigma$ minima
left. In the following we will call the minimum which is closest to
the true parameter values the {\it true}\/ minimum and all other minima
will be termed {\it wrong}\/ minima. Figure~\ref{Zusatz456} depicts the
evolution of the $3\sigma$ contours when the number $N_\indfrm{I}$
of interferometric measurements is further increased. It can be seen
here that the wrong minima are gradually vanishing. For
$N_\indfrm{I}=6$ the true minimum is the only minimum remaining and
therefore a definite time delay determination can be obtained. 
However, already for $N_\indrm{I}=4$ the {\it global}\/ minimum in the
$\chi^2$ function, which gives the best fit parameter values, 
corresponds to the true minimum.  
\begin{table}[t]
  \vspace{-2mm}
  \renewcommand{\baselinestretch}{0.8}
  \caption{\footnotesize Standard parameter values used for applying
the lightcurve reconstruction method to the simulations.}
  \label{recostandard}
  \begin{center}
  \vspace{-0mm}
\begin{tabular}{|l|c|c|} \hline
parameter&symbol&standard value \\ \hline\hline
relative source variability&$\nu$&2\% \\ \hline
number of interferometric observations&$N_\indrm{I}$&5 \\ \hline
interval between interferometric observations&$\Delta T_\indrm{I}$&5 \\ \hline
error of flux density ratio measurements&$\sig{m}/m$&0.5\% \\ \hline
error of the average combined flux density&$
\sig{\langle\tilde S_{1+2}\rangle}/\langle\tilde S_{1+2}\rangle$&2\% \\ \hline
\end{tabular}
  \vspace{-4mm}
  \end{center}
\end{table}
\begin{figure}[b]
  \vspace{-0mm}
    \leavevmode
    \epsfxsize=10cm
    \epsffile{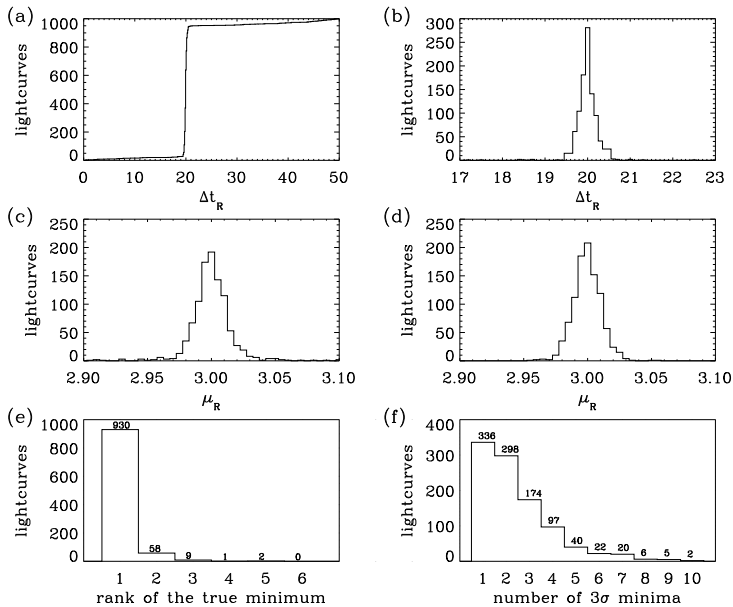}
  \vspace{-0mm}
\parbox[b]{5.8cm}{
  \renewcommand{\baselinestretch}{0.8}
  \caption{\footnotesize Plot {\bf(a)} shows the cumulative
distribution of $\dtrf$ values at the global minimum of the
$\chi^2$ function, whereas in {\bf(b)} the distribution of $\dtrf$
values at the true minimum is depicted. Accordingly plots {\bf(c)} and
{\bf(d)} show the distribution of the $\murf$ values at the global 
and at the true minimum. Diagram {\bf(e)} is a histogram of the
rank of the true minimum when ordering the minima according to
rising $\chi^2$ values. Finally, {\bf(f)} shows the distribution of
the number of $3\sigma$ minima appearing in the $\chi^2$ function.}   
  \label{ReconstSimul}
  \vspace{1mm}   }
\end{figure}

To test the lightcurve reconstruction method we again used
simulated lightcurves with parameters as shown in Table~\ref{lightstandard}.
Additional parameters which have to be specified for the lightcurve
reconstruction are displayed in Table~\ref{recostandard}. This table
also shows the ``standard values'' that we have adopted for these
parameters in the simulations. For simplicity the intervals between the
interferometric observations are assumed to be constant. These
observations are performed towards the end of the observing period $T$ when
the reconstruction is most accurate. Again we considered all minima in
the range $[0,50]$ as potential time delay values.\footnote{In
practice, of course, a plausible range of potential time
delay values can be given if a lens model is available. For B0218+357, e.g.,
the time delay is expected to be roughly between 8 and 20 days.}
Figure~\ref{ReconstSimul} summarizes the results of applying the
reconstruction to $1000$ simulated lightcurves with the parameter 
values of Tables~\ref{lightstandard} and \ref{recostandard}. 
Figure~\ref{ReconstSimul}a shows the cumulative distribution of the
$\dtr$ values at the global minimum of the $\chi^2$
function. From this plot it can be seen that for most of the simulated
lightcurves the location of the global minimum in fact agrees
with the value for the time delay which has been used for generating
them. To get an estimate of the accuracy achievable for the time delay
determination, Fig.~\ref{ReconstSimul}b depicts the distribution of the
$\dtr$ values at the true minimum. The narrow distribution
centered on the exact value of $\Delta t=20$ shows that a fairly high
precision can be expected with the standard parameter values used here.  
However, to obtain error ranges for individual realizations, $\chi^2$
statistics can be applied. This will be discussed in Sect.~\ref{minim}.
Plots c and d of Fig.~\ref{ReconstSimul} show the distribution of the
$\mur$ values at the global and at the true minimum,
respectively. For the rather small lightcurve variability of $\nu=2\%$
it is not surprising that the magnification ratio is well determined here. 
The two distributions are nearly identical, because also in the few
cases where the global minimum is a wrong minimum,
it is located roughly at the correct value for the magnification ratio
(cf. Fig.~\ref{Zusatz123}d). Figure~\ref{ReconstSimul}e shows the
distribution of the rank of the true minimum when ordering the
minima according to rising $\chi^2$ values. For $93\%$ of the
lightcurves the global minimum in the $\chi^2$ function
corresponds to the true minimum, and only for $\approx 1\%$ the
rank of the true minimum is higher than 2. Finally,
Fig.~\ref{ReconstSimul}f displays the distribution of the number of
$3\sigma$ minima which are appearing in the $\chi^2$ function. In the
majority of cases just one or a few minima are occurring, but there are
also rare cases with a large number of minima showing up.   
\begin{figure}[b]
  \vspace{-2mm}
    \leavevmode
    \epsfxsize=8cm
    \epsffile{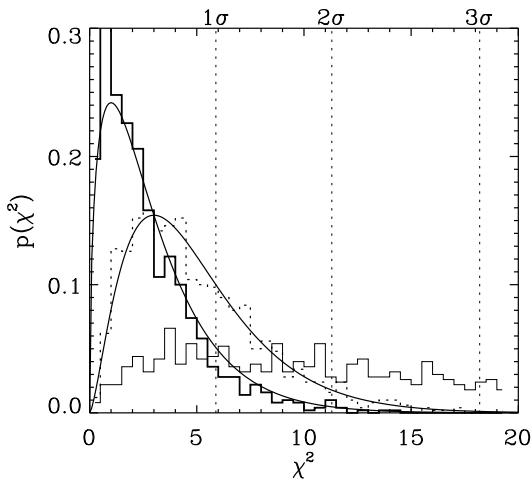}
  \vspace{-2mm}
\parbox[b]{7.8cm}{
  \renewcommand{\baselinestretch}{0.8}
  \caption{\footnotesize The thick, solid-line histogram shows the
distribution of $\chi^2$ values at the {\it true}\/ minimum overlaid
with the theoretical $\chi^2$ probability density distribution with
$5-2=3$ degrees of freedom. The dotted histogram represents the
distribution of $\chi^2$ values at the {\it exact}\/ parameter values
$\Delta t=20$ and $\mu=3$ overlaid with the theoretical $\chi^2$
distribution with $5$ degrees of freedom. The dotted vertical lines
indicate the $\chi^2$ limits for a given exclusion confidence
according to the theoretical distribution with $5$ degrees of
freedom. The thin, solid-line histogram displays the distribution of
$\chi^2$ values at the lowest {\it wrong}\/ minimum. (This
distribution is extending beyond the range of $\chi^2$ values covered
by this plot).}    
  \label{chiPlot}
  \vspace{4mm}   }
\end{figure}

In the following paragraph we will check if some statements given by $\chi^2$ 
statistics are applicable for this analysis. Let $N$ denote the number
of constraints used in a $\chi^2$ fit and $n$ be the number of fit
parameters. According to $\chi^2$ statistics the probability density
distribution for the $\chi^2$ value at the minimum of the $\chi^2$
function is given by a {\it $\chi^2$-distribution}\/ with $m:=N-n$
{\it degrees of freedom}\/, i.e.
\[
p_{\chi^2;m}(\chi^2)=\frac{1}{\Gamma(\frac{m}{2})\,2^{\frac{m}{2}}}
\chi^{\frac{m}{2}-1}e^{-\frac{\chi^2}{2}}~. 
\] 
This is valid if the errors are normally distributed and the model is
linear in its fit parameters. In our case, however, the model is
extremely non-linear in the fit parameter $\dtr$, and as we have
mentioned before this leads to the occurrence of additional wrong 
minima which cannot be described by $\chi^2$ statistics. As we have
seen in Fig.~\ref{ReconstSimul}e the wrong minima can be lower
than the true minima. Therefore the relation given above is
certainly not valid any more for the {\it global}\/ minimum of the
$\chi^2$ function. Nevertheless the $\chi^2$ values at the {\it
true}\/ minima still obey a $\chi^2$-distribution with $m$ degrees of
freedom. This is visualized by the thick, solid-line histogram in
Fig.~\ref{chiPlot} which shows the distribution of the $\chi^2$ values
at the true minimum for the simulated lightcurves and which
matches quite well with the overlaid theoretical $\chi^2$-distribution with
$5-2=3$ degrees of freedom as it is expected for $N_\indrm{I}=5$ constraints
and $2$ fit parameters. In addition, this figure reveals that the
distribution of the $\chi^2$ values at the {\it exact}\/ parameter
values is consistent with a $\chi^2$-distribution with $N_\indrm{I}=5$
degrees of freedom, again in agreement with the predictions of $\chi^2$
statistics. The thin, solid-line histogram in Fig.~\ref{chiPlot}
depicts the distribution of the $\chi^2$ values at the lowest {\it wrong}\/ 
minimum of the $\chi^2$ function which of course is not described by
$\chi^2$ statistics. The dotted vertical lines in the figure indicate
limiting $\chi^2$ confidence values for rejecting the hypothesis
that some given parameter values $\dtr$ and $\mur$ are the correct
ones. These exclusion confidence values have been derived from the
$\chi^2$-distribution with $N_\indrm{I}=5$ degrees of freedom for
$\chi^2$ at the exact parameter values. Corresponding definitions
already have been used for the $3\sigma$ contours in
Figs.~\ref{Zusatz123}d and \ref{Zusatz456} and for quantifying the
frequency of wrong minima in Fig.~\ref{ReconstSimul}f.
Actually this is a conservative limit for rejecting wrong
{\it minima}\/, because for the exclusion of the hypothesis that a given
minimum is the true one we would have to use the
$\chi^2$-distribution with $N_\indrm{I}-2$ degrees of freedom which is
shifted to smaller $\chi^2$ values. In Sect.~\ref{minim} we will show
with an example how to assign significance values to the statement
that one of several minima in the $\chi^2$ function is indeed the 
true minimum.
\subsection{Parameter dependence of the results}\label{parameters}
\begin{figure}[t]
  \vspace{-0mm}
    \leavevmode
    \epsfxsize=10cm
    \epsffile{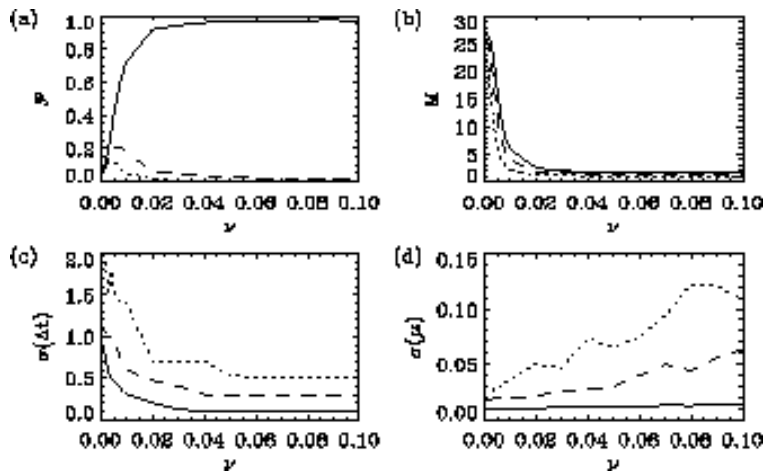}
  \vspace{-0mm}
\parbox[b]{5.8cm}{
  \renewcommand{\baselinestretch}{0.8}
  \caption{\footnotesize The dependence of the results of the
lightcurve reconstruction method on the variability
$\nu={\sigma_S}/{\langle\tilde S_{1+2}\rangle}$. Here the
noise parameter $\eta={\sigma_{\delta S}}/{\sigma_S}$ has been
adapted such that the single-dish observing errors $\sigma_{\delta S}$
remain constant.} 
  \label{variability}
  \vspace{3mm}   }
\end{figure}

\begin{figure}[b]
  \vspace{-0mm}
    \leavevmode
    \epsfxsize=10cm
    \epsffile{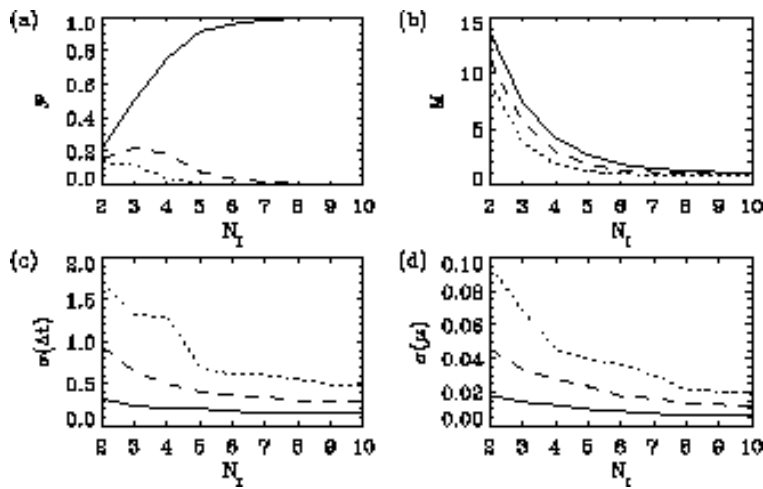}
  \vspace{-0mm}
\parbox[b]{5.8cm}{
  \renewcommand{\baselinestretch}{0.8}
  \caption{\footnotesize The dependence of the results of the
lightcurve reconstruction method on the number $N_\indfrm{I}$ of
interferometric observations.} 
  \label{intnumber}
  \vspace{3mm}   }
\end{figure}

\begin{figure}[t]
  \vspace{-0mm}
    \leavevmode
    \epsfxsize=10cm
    \epsffile{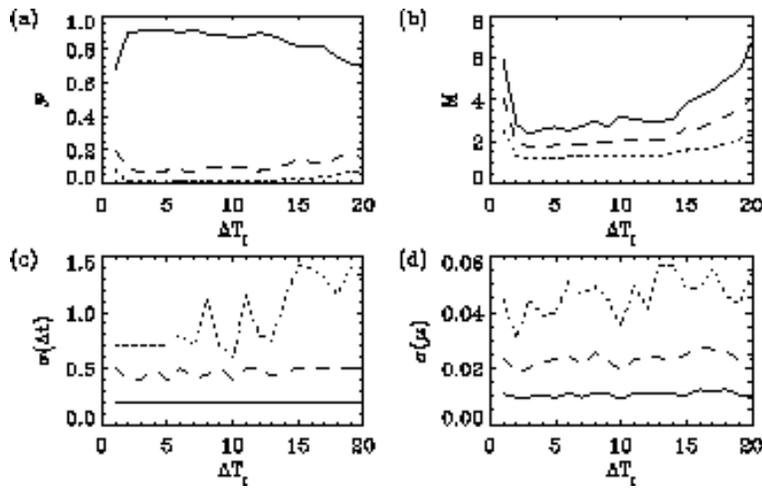}
  \vspace{-0mm}
\parbox[b]{5.8cm}{
  \renewcommand{\baselinestretch}{0.8}
  \caption{\footnotesize The dependence of the results of the
lightcurve reconstruction method on the time interval $\Delta T_\indfrm{I}$
between the interferometric observations.} 
  \label{intinterval}
  \vspace{3mm}   }
\end{figure}

In order to investigate the parameter dependence of the performance of the
lightcurve reconstruction method we introduce the following measures
to quantify the information which for the standard parameter values is
contained in Fig.~\ref{ReconstSimul}:  
\begin{itemize}
\item The probability ${\cal P}_i$, $i=1,2,3$ for the true
minimum being the global, the second lowest or the third lowest
minimum in the $\chi^2$ function (cf. Fig.~\ref{ReconstSimul}e).
\item The average number $M_i$, $i=1,2,3$ of $1\sigma$, $2\sigma$ and 
$3\sigma$ minima appearing in the $\chi^2$ function
(cf. Fig.~\ref{ReconstSimul}f).
\item Estimates $\sigma(\Delta t)$ and $\sigma(\mu)$ for the errors of the
position of the true minimum derived from the distributions of the
$\dtr$ and $\mur$ values at this minimum (cf. Fig.~\ref{ReconstSimul}b
and d). 
\end{itemize}
In the following we study the dependence of these quantities on the
parameters shown in Tables~\ref{lightstandard} and \ref{recostandard}.
Each time one parameter is varied and the others are kept fixed at the
standard values. 
\begin{figure}[b]
  \vspace{-4mm}
    \leavevmode
    \epsfxsize=10cm
    \epsffile{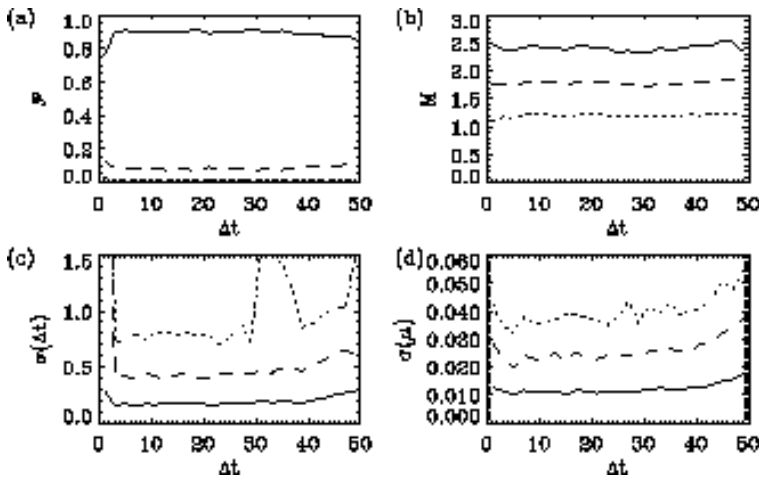}
  \vspace{-3mm}
\parbox[b]{5.8cm}{
  \renewcommand{\baselinestretch}{0.8}
  \caption{\footnotesize The dependence of the results of the
lightcurve reconstruction method on the time delay $\Delta t$. All
other parameters are kept fixed at the standard values given in
Tables~\protect\ref{lightstandard} and \protect\ref{recostandard}.
Plot {\bf(a)} shows the variation of ${\cal P}_1$ (solid line), 
${\cal P}_2$ (dashed line) and ${\cal P}_3$ (dotted line). Plot
{\bf(b)} depicts the quantities $M_3$ (solid line), $M_2$ (dashed
line) and $M_1$ (dotted line). Diagrams {\bf(c)} and {\bf(d)} show 
$1\sigma$ (solid line), $2\sigma$ (dashed line) and $3\sigma$ (dotted
line) error estimates for the time delay and the magnification
ratio. Note that the $3\sigma$ lines are poorly determined, because
they are calculated from the $\dtrf$ and $\murf$ values for a few
lightcurves only. (See the text for the definition of the quantities
displayed in this figure.)}     
  \label{timedelay}
  \vspace{-0mm}   }
\end{figure}
\paragraph{Time delay $\Delta t$:}
Figure~\ref{timedelay} presents the results for changing the value of
the time delay $\Delta t$ between $0$ and $50$. This corresponds to the
$\dtr$ range which is considered as the range of potential time delay
values and for which the $\chi^2$ function is calculated. 
The graphs indicate that apart from some boundary effects the results
are rather insensitive to varying $\Delta t$. They get slightly worse,
i.e. the fraction ${\cal P}_1$ is decreasing and the errors
$\sigma(\Delta t)$ and $\sigma(\mu)$ are increasing, for large time
delays $\Delta t$, because the errors of the reconstructed lightcurves
at the epoch of the interferometric observations increase with rising
$\dtr$ due to a smaller number of error-reducing iteration steps
needed for the reconstruction (cf. Figs.~\ref{RichtigFalsch} and
\ref{ReconstDemo}). Of course, the result also gets worse for very
short values of $\Delta t$ which are then becoming comparable to the
timescale of variability and the observing interval $\Delta T$.
\begin{figure}[t]
  \vspace{-4mm}
    \leavevmode
    \epsfxsize=10cm
    \epsffile{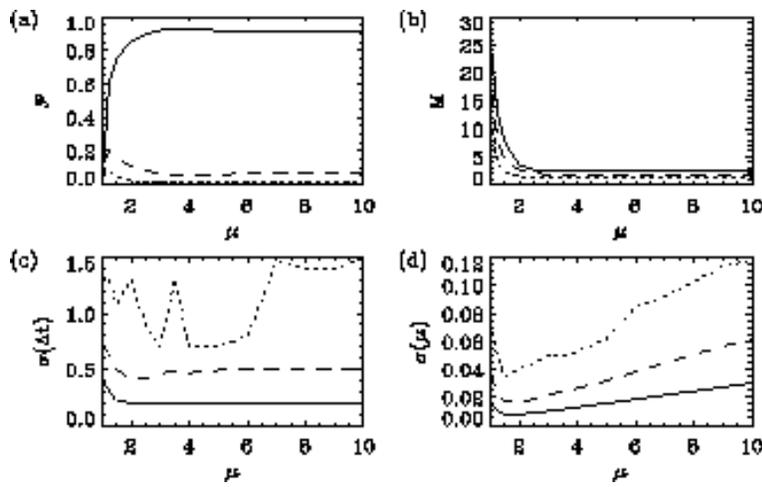}
  \vspace{-4mm}
\parbox[b]{5.8cm}{
  \renewcommand{\baselinestretch}{0.8}
  \caption{\footnotesize The dependence of the results of the
lightcurve reconstruction method on the magnification ratio $\mu$. See
the caption of Fig.~\protect\ref{timedelay} for the explanation of the various
lines in these plots.}   
  \label{magratio}
  \vspace{3mm}   }
\end{figure}
\paragraph{Magnification ratio $\mu$:}
Figure~\ref{magratio} verifies the statement about the magnification
ratio which already has been given in Sect.~\ref{reconstruction}: 
The method completely fails for $\mu\rightarrow 1$ because then the
errors due to the arbitrary initial guess do not decrease during the
reconstruction process. In the diagrams the breakdown of the method
for $\mu\rightarrow 1$ is reflected by the steep decrease of ${\cal P}_1$ 
and the steep increase of the number $M_i$ of minima appearing in the
$\chi^2$ function. For $\mu\rightarrow\infty$ the vaules ${\cal P}_1$, $M_i$
and $\sigma(\Delta t)$ are remaining virtually constant. Note,
however, that in this figure the relative accuracy of the flux density
ratio measurements has been kept constant. For rising magnification
ratio it will become increasingly difficult to measure the then also
rising flux density ratios with the desired accuracy, and this will
limit the applicability of the method for large $\mu$.
\begin{figure}[b]
  \vspace{-3mm}
    \leavevmode
    \epsfxsize=10cm
    \epsffile{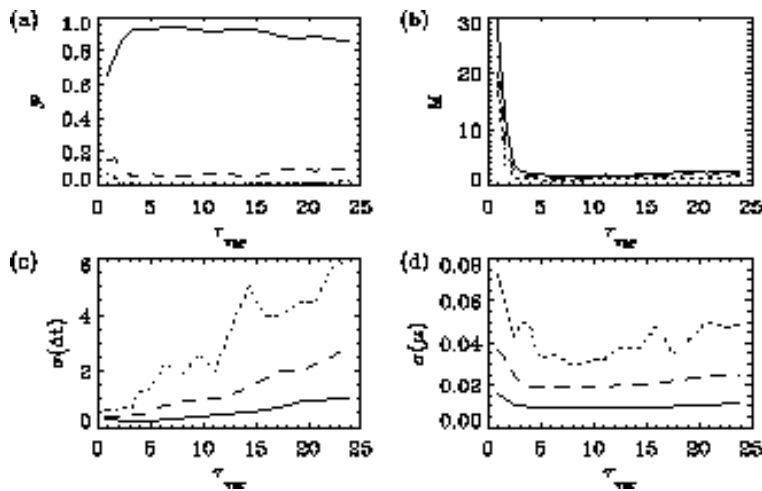}
  \vspace{-3mm}
\parbox[b]{5.8cm}{
  \renewcommand{\baselinestretch}{0.8}
  \caption{\footnotesize The dependence of the results of the
lightcurve reconstruction method on the timescale of variability
$\tau_\indfrm{var}$.}    
  \label{timescale}
  \vspace{3mm}   }
\end{figure}
\paragraph{Timescale of variability $\tau_\indrm{var}$:}
Figure~\ref{timescale}, showing the dependence of the results on the
timescale of variability, mainly reveals two effects. For very short 
timescales $\tau_\indrm{var}\lesssim 2$ the errors introduced by
interpolating the observed lightcurve between consecutive data points
become large. This leads to an increase in the number of unwanted
wrong minima and to a decrease of the fraction ${\cal P}_1$.   
For slow variation, on the other hand, the number of minima roughly stays
constant and ${\cal P}_1$ is only slightly decreasing, at least in the
range of timescales plotted here. However, the error $\sigma(\Delta t)$ 
in the location of the true minimum is rising then. This means
that the minima in the $\chi^2$ function are becoming broader and that
the accuracy of the time delay determination is reduced.
\paragraph{Variability $\nu$:}
A very important parameter is, of course, the relative variability of
the source. In Fig.~\ref{variability} not only
$\nu={\sigma_S}/{\langle\tilde S_{1+2}\rangle}$ has been varied,
but also the noise parameter $\eta={\sigma_{\delta S}}/{\sigma_S}$,
which also includes the dispersion $\sigma_S$ of the combined lightcurve, has 
been adapted in such a way that the absolute errors $\sigma_{\delta S}$
of the single-dish lightcurve measurements remain constant.
Naturally for $\nu=0$ no time delay determination is possible, but for
increasing variability the results improve quite quickly, and ${\cal P}_1$ 
nearly reaches the ideal value of $1.0$ for $\nu\gtrsim 0.04$. 
The simulations have not been extended to variabilities larger
than $10\%$ because then the minima in the $\chi^2$ function are
becoming very steep and narrow, which makes it more difficult to
find them numerically.      
\paragraph{Number of interferometric observations $N_\indrm{I}$:} 
Figure~\ref{intnumber} shows the evolution of the results when
increasing the number of additional interferometric observations which
are used as constraints for the $\chi^2$ fit. From these diagrams it
can be seen how the fraction of lightcurves for which the true 
minimum is the global minimum of the $\chi^2$ function rises and
the number of minima and the errors in the location of the
true minimum decrease. For eight flux density ratio measurements,
the fraction ${\cal P}_1$ has nearly reached $100\%$ and typically
there is just one $3\sigma$ minimum remaining. 
\paragraph{Interval between interferometric observations 
$\Delta T_\indrm{I}$:} 
For simplicity this study has been restricted to constant time
intervals between the interferometric observations. These are performed
towards the end of the observing period when the reconstruction is most
accurate (cf. Fig.~\ref{FlussZeit}). Figure~\ref{intinterval} shows
that the standard value $\Delta T_\indrm{I}=5$ is almost ideal for the
standard parameter set. The results get worse for small 
$\Delta T_\indrm{I}$ which are comparable to the timescale of variability,
because then consecutive flux density ratio measurements do not
provide independent constraints. Therefore larger values of
$\Delta T_\indrm{I}$ are adequate for more slowly varying
lightcurves. Very large values for $\Delta T_\indrm{I}$, however, are
disfavourable, because then the first interferometric observations
would have to be placed early on during the observing period when the
errors of the reconstruction are still large.    
\begin{figure}[b]
  \vspace{-0mm}
    \leavevmode
    \epsfxsize=10cm
    \epsffile{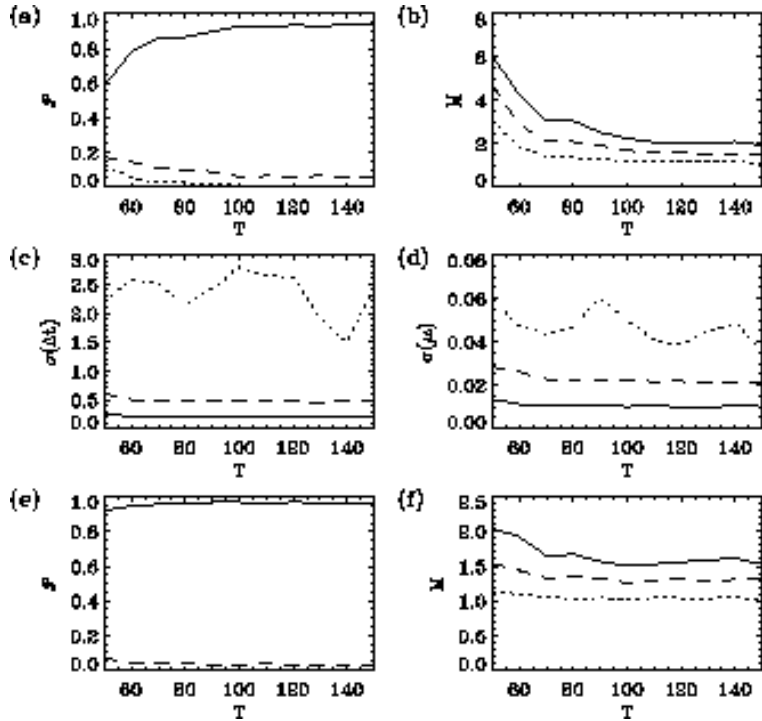}
  \vspace{-0mm}
\parbox[b]{5.8cm}{
  \renewcommand{\baselinestretch}{0.8}
  \caption{\footnotesize The dependence of the results of the
lightcurve reconstruction method on the observing period $T$. Plots
{\bf(a)}-{\bf(d)} are of the same structure as in the previous
figures. Diagrams {\bf(e)} and {\bf(f)} show the results for ${\cal
P}_i$ and $M_i$ for the same set of parameters as in {\bf(a)} and
{\bf(b)}. However, the interval considered as the range of potential
time delay values has been reduced here from $[0,50]$ to $[0,30]$.} 
  \label{obsperiod}
  \vspace{5mm}   }
\end{figure}
\paragraph{Observing period $T$:}
In contrast to the autocorrelation method (cf. Fig.~\ref{autoTmue}a) the
dependence of the results on the observing period $T$ is rather
indirect for the reconstruction method. The need for a sufficiently
long observing period arises from the fact that the uncertainty due to
the arbitrary initial guess gradually decreases during the
reconstruction process. Eventually, the errors of the reconstructed
lightcurves will be dominated by the observational errors of the
single-dish measurements and will therefore roughly stay constant with time.
This behaviour is reflected by the diagrams a to d of
Fig.~\ref{obsperiod}. For small $T$ the results improve when extending
the observation period, because then the interferometric measurements are still
performed at epochs for which the reconstruction is affected by the
initial errors. But since these have already dropped out at the respective
epochs for $T\gtrsim100$, prolonging the observation period beyond that
value does not improve the results any longer.
  
\begin{figure}[b]
  \vspace{-0mm}
    \leavevmode
    \epsfxsize=10cm
    \epsffile{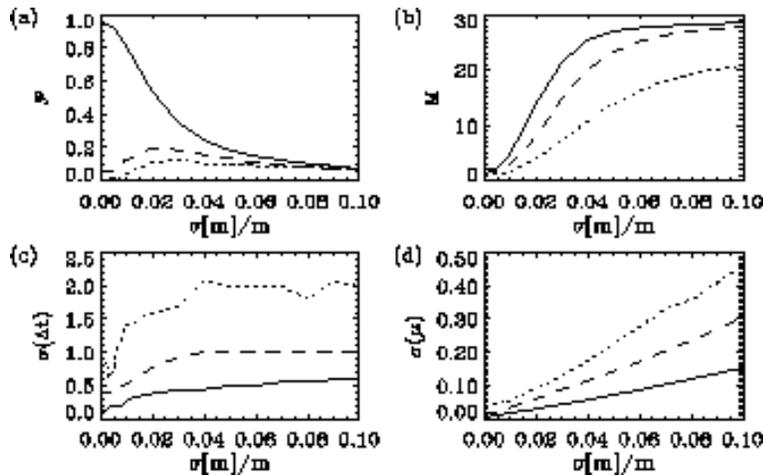}
  \vspace{-0mm}
\parbox[b]{5.8cm}{
  \renewcommand{\baselinestretch}{0.8}
  \caption{\footnotesize The dependence of the results of the
lightcurve reconstruction method on the relative error of the
interferometric flux density ratio measurements.} 
  \label{ratioerror}
  \vspace{3mm}   }
\end{figure}

Figure~\ref{obsperiod}e and f show the effects of reducing the interval of
potential time delay values, for which the reconstruction is done and
for which the $\chi^2$ function is calculated, from $[0,50]$ to $[0,30]$. Of
course, this significantly improves the results, because then the 
wrong minima which are located in the interval $]30,50]$ are
excluded from the analysis. In addition, the worsening of the results
for $T\lesssim100$ is not so serious in this case, because these
problems preferentially arise due to the large $\dtr$ values for which
the errors of the reconstruction are decreasing slower, and just these
$\dtr$ values have been excluded here.    
\begin{figure}[b]
  \vspace{-0mm}
    \leavevmode
    \epsfxsize=8cm
    \epsffile{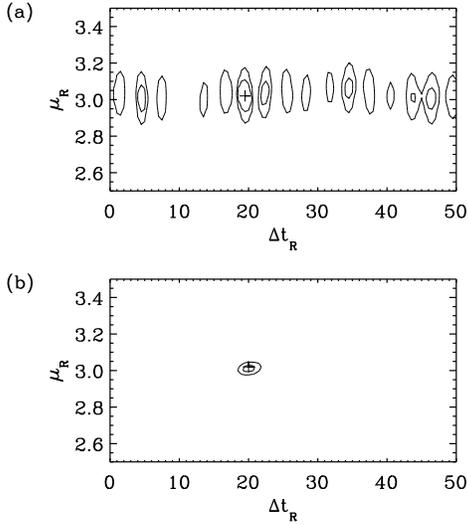}
  \vspace{-0mm}
\parbox[b]{7.8cm}{
  \renewcommand{\baselinestretch}{0.8}
  \caption{\footnotesize {\bf(a)} This diagram shows the $1\sigma$ and
$3\sigma$ contours of the $\chi^2$ function for an individual
lightcurve realization with the standard parameter values, except
that the single-dish observing interval has been changed to $\Delta T=3$.
{\bf(b)} For this example the observing interval is $\Delta T=3$ as
well, but in addition to that the timescale of variability has been
increased to $\tau_\indfrm{var}=15.9$.} 
  \label{obsinterval}
  \vspace{3mm}   }
\end{figure}
\paragraph{Observing interval $\Delta T$:} 
The effects of changing the observing interval $\Delta T$ for the
single-dish measurements are conveniently demonstrated by inspecting
the $\chi^2$ contours for an individual lightcurve realization.
We have mentioned before that for calculating the reconstruction with
$\dtr$ values which are not integer multiples of $\Delta T$, the observed
combined lightcurve has to be interpolated which in general causes
additional errors for the reconstructed lightcurves. This leads to the
effect that integer multiples of $\Delta T$ can be excluded as values
for the time delay with a higher confidence than non-integer multiple
values and that wrong minima are preferentially appearing at
the latter ones. This is drastically demonstrated by the example shown
in Fig.~\ref{obsinterval}a. Here the observing interval has been
changed to $\Delta T=3$ and a variety of wrong minima are showing
up. Note, however, that the global minimum still corresponds
to the true minimum. In practice the effect described here
will be alleviated to some extent by irregular observing intervals.
The example plotted in Fig.~\ref{obsinterval}b shows that the
adequate choice of the observing interval sensitively depends on the
timescale of variability. Despite using the same observing interval of
$\Delta T=3$ as in Fig.~\ref{obsinterval}a only one minimum is
appearing here and the time delay can unambiguously be determined, 
because in this case the variability timescale $\tau_\indrm{var}=15.9$ is
very much longer and therefore the interpolation errors are not so serious.
\begin{figure}[t]
  \vspace{-0mm}
    \leavevmode
    \epsfxsize=10cm
    \epsffile{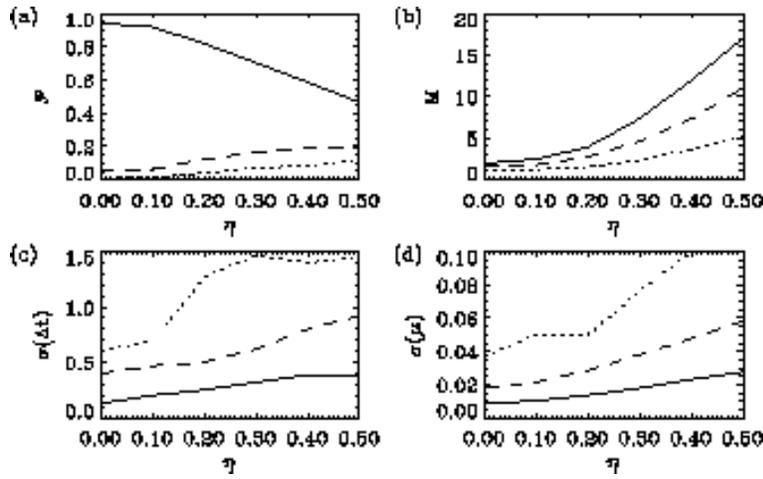}
  \vspace{-0mm}
\parbox[b]{5.8cm}{
  \renewcommand{\baselinestretch}{0.8}
  \caption{\footnotesize The dependence of the results of the
lightcurve reconstruction method on the errors of the single-dish flux
density measurements, characterized by the parameter 
$\eta=\sigma_{\delta S}/\sigma_S$.} 
  \label{obserror}
  \vspace{3mm}   }
\end{figure}
\paragraph{Observing erros $\eta$:}
From Fig.~\ref{obserror} it can be seen that fairly accurate
single-dish flux density measurements are required for the method to work.
For the standard parameter values the rms errors $\sigma_{\delta S}$
should not exceed  $10\%$ of the dispersion $\sigma_S$ of the
lightcurve itself ($\eta=\sigma_{\delta S}/\sigma_S=0.1$).  
The reason for the sensitivity to $\eta$ is, of course, that errors in
the observed total lightcurve directly translate into errors of the
reconstructed individual lightcurves and flux density ratios. In
particular, it is desirable to determine the total flux density to a
very good accuracy at the epochs of the interferometric observations,
because the errors of these values contribute to the errors of the
predicted flux density ratios to a large extent (cf. Eqs.~(\ref{recflux}) 
and (\ref{flussfehler})).   
\paragraph{Flux density ratio erros $\sig{m}/m$:}
A crucial parameter is the accuracy of the additional constraints
which are provided by the interferometric observations. 
Figure~\ref{ratioerror} depicts the dependence of the results on the
relative error $\sig{m}/m$ of the flux density ratio measurements.
For the standard parameter values this error should not exceed
$\approx1\%$.

\bigskip
To complete the discussion of the parameter dependence we mention here
that the effects of increasing the error $\sig{\langle\tilde S_{1+2}\rangle}$
of the average combined flux density, which enters in
Eq.~(\ref{flussfehler}), are negligible even for high values.
Furthermore, changing the shape of the power spectrum which is used
for generating the simulated lightcurves does not affect the results
significantly.  
\subsection{$\Delta t$ determination for individual realizations}\label{minim} 
\begin{figure}[h]
  \vspace{0mm}
    \leavevmode
    \epsfxsize=8cm
    \epsffile{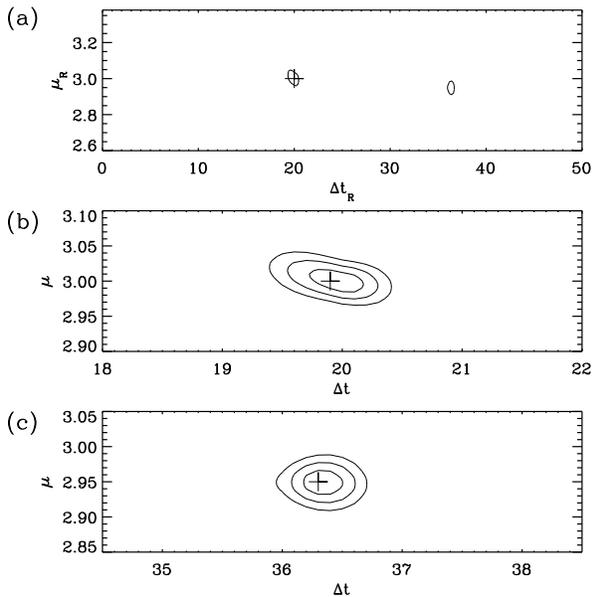}
  \vspace{0mm}
\parbox[b]{7.8cm}{
  \renewcommand{\baselinestretch}{0.8}
  \caption{\footnotesize {\bf(a)} The $3\sigma$ contours (as
defined in Sect.~\protect\ref{chisqu}) for one lightcurve realization with the
standard parameter values. The cross marks the global minimum. Plot
{\bf(b)} shows the $3\sigma$, $2\sigma$ and $1\sigma$ confidence
regions for $\Delta t$ and $\mu$ under the assumption 
that the minimum at $\dtrf=19.9$ is indeed the true minimum.  
Accordingly {\bf(c)} shows the confidence regions for assuming
that the minimum at $\dtrf=36.3$ is the true minimum.
The limiting $\chi^2$ values for these confidence regions were derived
from the $\chi^2$-distribution with 2 degrees of freedom for
$\Delta\chi^2=\chi^2_\indfrm{exact}-\chi^2_\indfrm{true}$ 
(see Fig.~\protect\ref{deltachi}). Thus, diagrams {\bf(b)} and
{\bf(c)} show the contours of
$\chi^2=\chi^2_\indfrm{true}+\Delta\chi^2$ with $\Delta\chi^2$ taken
from the vertical lines in Fig.~\protect\ref{deltachi} and
$\chi^2_\indfrm{true}$ taken as the $\chi^2$ value at the respective 
minimum.}   
  \label{chiexample}
  \vspace{3mm}   }
\end{figure}

Figure~\ref{chiexample}a shows the $\chi^2$ contours for one example
lightcurve generated with the standard parameter values
(Tables~\ref{lightstandard} and \ref{recostandard}). In this case two
minima appear with $\chi^2$ values low enough to be considered as candidates
for the true minimum. Ideally one would like to include enough
interferometric measurements so that only one sufficiently deep
minimum is remaining. But here we have chosen this example
in order to illustrate how we can determine probabilities for particular
minima to be the true minimum in such cases and how error ranges
for the time delay and the magnification ratio can be obtained. 
The former will not be possible without making use of the simulations,
but concerning the latter point we can again apply $\chi^2$ statistics.

\begin{figure}[t]
  \vspace{-3mm}
    \leavevmode
    \epsfxsize=8cm
    \epsffile{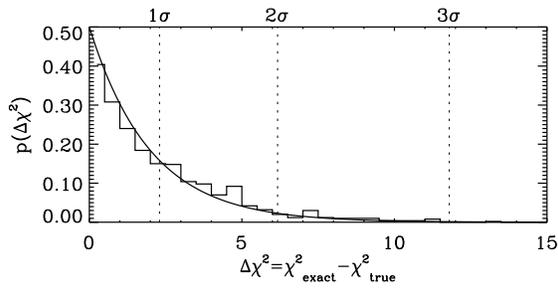}
  \vspace{-3mm}
\parbox[b]{7.8cm}{
  \renewcommand{\baselinestretch}{0.8}
  \caption{\footnotesize The histogram shows the distribution of
$\Delta\chi^2=\chi^2_\indfrm{exact}-\chi^2_\indfrm{true}$ determined from the
simulated lightcurves (with standard parameters) that have already
been used for generating Figs.~\protect\ref{ReconstSimul} and
\protect\ref{chiPlot}. This distribution is consistent with the
theoretical $\chi^2$-distribution with 2 degrees of freedom which is
plotted in the diagram as well. According to this distribution the
dotted vertical lines indicate the limiting $\Delta\chi^2$ values for
a given exclusion confidence.}     
  \label{deltachi}
  \vspace{3mm}   }
\end{figure}

Figure~\ref{deltachi} verifies that the distribution of the difference
$\Delta\chi^2=\chi^2_\indrm{exact}-\chi^2_\indrm{true}$ between the
$\chi^2$ values at the exact parameter values ($\Delta t$, $\mu$) and at the
true minimum is following a $\chi^2$-distribution with 2
degrees of freedom, as it is theoretically expected for two fit
parameters ($\dtr$ and $\mur$). Assuming that a particular minimum is
indeed the true minimum, we can use the limiting $\Delta\chi^2$
values indicated in Fig.~\ref{deltachi} to obtain confidence regions
for the time delay $\Delta t$ and the magnification ratio $\mu$.  
This has been done in plots c and d of Fig.~\ref{chiexample} for the
minima at $\dtr=19.9$ and $\dtr=36.3$, respectively. Note
that for determining confidence intervals for the time delay $\Delta t$
alone, irrespective of the magnification ratio, the
$\chi^2$-distribution with 1 degree of freedom has to be used for
$\Delta\chi^2$.  

\begin{figure}[b]
  \vspace{-10mm}
    \leavevmode
    \epsfxsize=10cm
    \epsffile{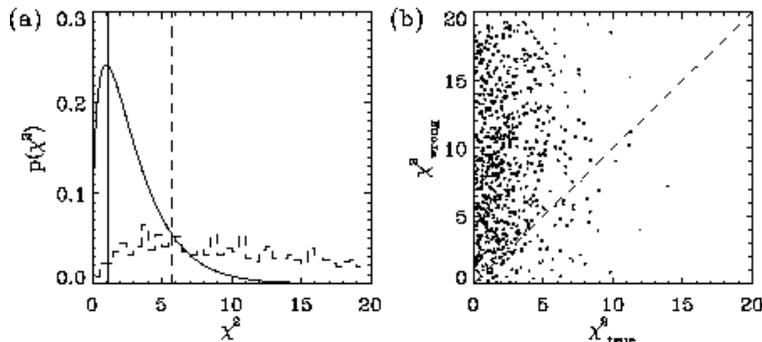}
  \vspace{-2mm}
\parbox[b]{5.8cm}{
  \renewcommand{\baselinestretch}{0.8}
  \vspace{-0mm}   
  \caption{\footnotesize {\bf(a)} This plot again shows the distributions of
the $\chi^2$ values at the true minimum, represented by the
theoretical curve (solid line), and at the lowest wrong
minimum (dashed line), determined from the simulations 
(cf. Fig.~\protect\ref{chiPlot}). The thick vertical lines indicate
the $\chi^2$ values at the two minima of the $\chi^2$ function shown
in Fig.~\protect\ref{chiexample} (solid line for $\dtrf=19.9$,
dashed line for $\dtrf=36.3$). {\bf(b)} In this scatter plot the
$\chi^2$ values at the lowest wrong minimum are plotted
against those at the true minimum for each realization of the
simulated lightcurves.}    
  \label{chiProb}
  \vspace{-2mm}   }
\end{figure}

In order to quantify the probability for the respective minima to be
the true minimum we distinguish the following two cases:
\begin{itemize}
\item[{\bf A}:]{The minimum at $\dtr=19.9$ is the true
minimum. Then the minimum at $\dtr=36.3$ necessarily has to be the
lowest wrong minimum.}
\item[{\bf B}:]{The minimum at $\dtr=36.3$ is the true
minimum. This means that the minimum at $\dtr=19.9$ is the
lowest wrong minimum.}
\end{itemize}
Here we are neglecting the possibility that both of these minima
are wrong, because this would imply that the true
minimum is at a $\chi^2$ value beyond the $3\sigma$ limit and thus very
far in the exponentially decreasing tail of the probability density
distribution (cf. Fig.~\ref{chiPlot})\footnote{Note that the $3\sigma$
limit which has been used for the contours in Fig.~\protect\ref{chiexample}a
and which is indicated by the vertical line in
Fig.~\protect\ref{chiPlot} is derived from the $\chi^2$-distribution
with $N_\indfrm{I}=5$ degrees of freedom and therefore is conservative
for the purpose of excluding wrong minima. To do that the
distribution with $N_\indfrm{I}-2=3$ degrees of freedom would be
adequate. See the discussion at the end of Sect.~\protect\ref{chisqu}.}.
In Fig.~\ref{chiProb}a we again show the probability density distribution 
$p_\indrm{true}(\chi^2)$ for the true minimum, represented by
the theoretical $\chi^2$-distribution with 3 degrees of freedom, and
the probability density distribution $p_\indrm{wrong}(\chi^2)$ for the
lowest wrong minimum as determined from the simulations. In
the following we denote the $\chi^2$ values at the $\dtr=19.9$
minimum with $\chi^2_\indrm{a}$ and those at the $\dtr=36.3$ minimum
with $\chi^2_\indrm{b}$. In the figure these values are marked with
vertical lines, the solid line for $\chi^2_\indrm{a}\approx1$ and
the dashed line for $\chi^2_\indrm{b}\approx6$. In case {\bf A}, i.e.
$\dtr=19.9$ being the true minimum, the value for
$\chi^2_\indrm{a}$ is drawn from the distribution $p_\indrm{true}(\chi^2)$ 
and therefore $\chi^2_\indrm{b}$ is drawn from $p_\indrm{wrong}(\chi^2)$,
whereas for case {\bf B} it is vice versa. In Fig.~\ref{chiProb}b the
values $\chi^2_\indrm{wrong}$ at the lowest wrong minimum have
been plotted against the value $\chi^2_\indrm{true}$ at the true
minimum for the individual realizations of the lightcurve
simulations. Since this scatter plot shows no correlation between
these $\chi^2$ values, we can treat $p_\indrm{true}(\chi^2)$ and
$p_\indrm{wrong}(\chi^2)$ as independent distributions.
This allows us to assign probabilities to the cases {\bf A} and {\bf B}:  
\[
{\cal P}_\indrm{A}=
\frac{p_\indrm{true}(\chi^2_\indrm{a})\,p_\indrm{wrong}(\chi^2_\indrm{b})}
{p_\indrm{true}(\chi^2_\indrm{a})\,p_\indrm{wrong}(\chi^2_\indrm{b})+
p_\indrm{true}(\chi^2_\indrm{b})\,p_\indrm{wrong}(\chi^2_\indrm{a})}
\approx91\%~,
\]
\[
{\cal P}_\indrm{B}=
\frac{p_\indrm{true}(\chi^2_\indrm{b})\,p_\indrm{wrong}(\chi^2_\indrm{a})}
{p_\indrm{true}(\chi^2_\indrm{a})\,p_\indrm{wrong}(\chi^2_\indrm{b})+
p_\indrm{true}(\chi^2_\indrm{b})\,p_\indrm{wrong}(\chi^2_\indrm{a})}
\approx9\%~.
\]
Concerning the determination of the time delay and the magnification
ratio we can therefore make the following statements. With a
probability of $91\%$ the minimum at $\dtr=19.9$ is the true 
minimum. Then confidence regions for the time delay $\Delta t$
and the magnification ratio $\mu$ are given by Fig.~\ref{chiexample}b.
The probability for $\dtr=36.3$ being the true minimum is $9\%$.
In this case the confidence regions for $\Delta t$ and $\mu$ are given
by Fig.~\ref{chiexample}c. In view of the much higher probability of
case {\bf A}, which is indeed the true one, this result could also be
phrased as: The error level for restricting the time delay to be 
$\Delta t=19.9$ with confidence regions as shown in Fig.~\ref{chiexample}b
is $9\%$.

At this point a few cautionary remarks should be added. The probabilities
derived here rely on the distribution of the wrong minima which
cannot be described by $\chi^2$ statistics and therefore has to be determined
from the simulations. Hence, in applications to real observations it is
important to generate a set of simulated lightcurves which
satisfactorily reflects the variability characteristics of the observed
lightcurve. Determining the confidence regions in
Fig.~\ref{chiexample}b and c, however, solely relies on $\chi^2$
statistics and does not require any simulations. Of course, for any
$\chi^2$ fit it is essential to use error estimates which are really
reflecting the statistical uncertainties of the measurements and not,
e.g., the observer's predilection for giving conservative limits.  
\begin{figure}[b]
  \vspace{-3mm}
    \leavevmode
    \epsfxsize=10cm
    \epsffile{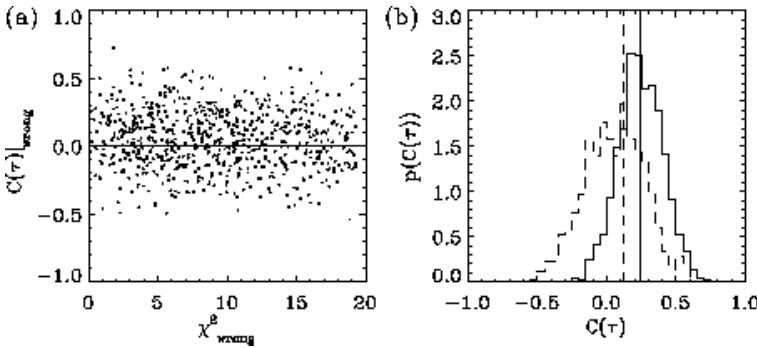}
  \vspace{-3mm}
\parbox[b]{5.8cm}{
  \renewcommand{\baselinestretch}{0.8}
  \caption{\footnotesize {\bf(a)} In this diagram the values of the
autocorrelation function for lags $\tau$ corresponding to the $\dtrf$
value of the lowest wrong minimum are plotted against the
$\chi^2$ value at that minimum. {\bf(b)} This plot shows the
distribution of the $C(\tau)$ values for $\tau$ corresponding to the 
true minimum (solid line) and to the lowest wrong
minimum (dashed line). In analogy to Fig.~\protect\ref{chiProb}a the
values at the minima of Fig.~\protect\ref{chiexample} have been
indicated by thick vertical lines.}
  \label{corrProb}
  \vspace{1mm}   }
\end{figure}

To conclude this section we demonstrate how the results can be
improved by additionally taking into account the information contained
in the autocorrelation function. In Fig.~\ref{corrProb}a the value 
$C(\tau)|_\indrm{wrong}$ of the autocorrelation function for the lag $\tau$
corresponding to the time delay value $\dtr$ of the lowest wrong  
minimum has been plotted against the $\chi^2$ value at that minimum
for the individual lightcurve realizations. The scatter plot indicates
that there is no correlation between these two quantities and that
on average the value $C(\tau)|_\indrm{wrong}$ is zero, i.e. there is
no enhanced autocorrelation for $\tau$ values corresponding to the
$\dtr$ values of wrong minima. This implies that the
lightcurve reconstruction method and the autocorrelation method are
completely independent techniques for determining the time delay.
Figure~\ref{corrProb}b depicts the distributions of the autocorrelation
values for lags $\tau$ corresponding to the $\dtr$ value of the 
true minimum (solid line) and of the lowest wrong
minimum (dashed line). The latter distribution is centered around
$C(\tau)\approx0$, whereas the former is shifted to higher
autocorrelation values. The reason for this is, of course, that the
true minimum is roughly located at the correct time delay
$\Delta t$ and so the effect discussed in Sect.~\ref{autocorrelation}
leads to an enhanced autocorrelation. From the large overlap of these
distributions and from the small separation of the vertical lines in
this plot, which in analogy to Fig.~\ref{chiProb}a denote the values
for the two minima of the example case, it is clear that the
autocorrelation function cannot improve the results drastically. Nevertheless,
including this additional information in a similar analysis as above leads to
probability values of ${\cal P}_\indrm{A}\approx95\%$ and 
${\cal P}_\indrm{B}\approx5\%$ and thus the error level for tying
oneself down to $\Delta t=19.9$ as the value for the time delay
could be reduced from $9\%$ to $5\%$.

%
% chapter4
%
\section{Discussion and conclusions}\label{discussion}
In the first part of this paper we have shown that an analysis of the
autocorrelation function of the total lightcurve of gravitational lens
systems can in principle be used to determine its time delay.
In practice, however, a fairly long observing time is required to make
sure that the {\it measured}\/ autocorrelation function reliably
reflects the lightcurve's underlying statistical properties.
In addition, the characteristic time scale for the source variability
has to be considerably shorter then the time delay, and the
magnification ratio should be close to unity in order for the
autocorrelation function to develop distinct time delay features.

It is interesting to note that it has been proposed by Press (1996)
that in certain cicumstances time autocorrelation effects on
unresolved, time-varying objects could be used to establish otherwise
undetectable (micro-) lens systems. In view of the results
shown in Sect.~\ref{autocorrelation} a detection of the time delay
effect via an analysis of the autocorrelation function seems to
require a delicate fine tuning between the involved time scales,
i.e. the observing interval, the time scale of variability, the time
delay, the observing period and possibly the ``life time'' of the
lensing configuration. However, using different approaches might be
more promising, especially for non-gaussian random processes.
Consider, as an extreme example, an intrinsic source lightcurve
consisting of a characteristic series of sharp peaks. In such a
``burst-scenario'' a time delay could be easily determined from
the repetition of the features which is then showing up in the observed
lightcurve of the lens system. 

In Sect.~\ref{lightcurve} we introduced the ``lightcurve
reconstruction method''. We demonstrated that it is possible to reconstruct
the individual lightcurves of a gravitational lens system from the
observed total lightcurve by assuming values for the time delay and
the magnification ratio. The reconstructed lightcurves are most
accurate at the end of the observing period, because errors introduced
by an arbitrary initial guess are gradually decreasing during the
reconstruction process. However, this is not true for a magnification
ratio close to unity and the method is therefore not applicable in such cases. 
In general it will be impossible without additional information to
single out the true time delay from all possible reconstructions with
potential values for the time delay and the magnification ratio. (A
counterexample for which this might indeed be feasible by assuming
some reasonable shape for the correctly reconstructed lightcurves is 
the highly non-gaussian burst-scenario outlined above.) However,
interferometric  measurements of the flux density ratio (at the same
frequency as the total flux density monitoring!) provide an excellent
means for checking the consistency of the various reconstructions and
eventually determining the time delay unambiguously.  

With simulations we investigated the dependence of the method on various
parameters. Naturally the variability of the lightcurve turns out to be
an essential parameter -- as it is the case for any time delay
determination technique. In addition, a fairly high accuracy for the
total flux density monitoring observations as well as for the additional
interferometric flux density ratio measurements is desirable. In
contrast to the autocorrelation method the lightcurve reconstruction
also works for rather slowly varying lightcurves, although increasing
the time scale of variability leads to larger confidence intervals for
the time delay. (That lightcurves with fairly slow variability still
contain information about the time delay can also be seen from van
Ommen et al.'s (1995) analysis of the lightcurve of PKS1830-211, a
lens system with two compact images and an Einstein ring. They use
simple linear fits to the slowly varying total lightcurve and combine
these with interferometric flux density ratio measurements to get an
estimate for the time delay. However, the confidence intervals for
this value are still quite large.) Of course, the simulations that we
have performed just show the general trends and cannot cover the
parameter space completely. For a given lens system and observation
programme some unfavourable parameter values could be compensated by
another parameter being much more adequate and thus leading to
equivalently good results. 

In principle the method described here could be used for radio as well
as for optical lightcurves. However, there are several arguments against
optical wavelengths. If the time scale of variability is short, a small
observing interval is essential to sample the lightcurve accurately
and a larger gap due to bad weather would have drastic effects on the
quality of the reconstruction. In addition, microlensing effects which
cannot properly be acounted for in this method can be important in
the optical. However, if the microlensing time scale is considerably
larger than the time delay and the observing period, this will only
result in a change of the effective magnification ratio and the method
can still be applied. Finally, a very important reason to perform 
such observations at radio frequencies is that the most promising lens
systems for determining the Hubble constant consist of double images of
radio-loud sources and an associated ring of extended emission. 
From these Einstein rings the system B0218+357 mentioned in the
introduction probably is the best candidate at the moment. The time
delay for this system is expected to be roughly between 8 and 20 days
which is reflecting our current knowledge (or rather ignorance) of the
Hubble constant and uncertainties in the lens model. (A detailed model
for this system which takes into account the information provided
by the Einstein ring still has to be made.) The preliminary value of
$\Delta t=12\pm3$ for the time delay obtained by Corbett et al. (1996)
is consistent with this range. Reliable estimates for the time scale and the
amplitude of the source variability are required in order to design an
optimal observing strategy and to specify the number of
interferometric observations needed for a definite determination of
the time delay by using the lightcurve reconstruction method. However,
combining a single-dish monitoring programme of two or three months
with five to ten additional interferometric constraints should be 
adequate to determine the time delay reliably. 

For the system PKS1830-211 the discovery of molecular absorption
lines by Wiklind \& Combes (1996) opens an interesting new possibility to
determine the time delay from unresolved single-dish monitoring
observations only. Apparently only one of the two source images is covered
by a molecular cloud, and this allows one to determine the flux
density ratio from molecular spectroscopy. Combining this with the
total flux density it is possible to calculate the individual
lightcurves and apply a cross-correlation analysis in order to obtain
the time delay. However, the lightcurve reconstruction method could be
used in this case as well, with the advantage of a much cleaner error
treatment, because it uses the direct observables total flux density
and flux density ratio.

%
% acknowledgements
%

\bigskip
\noindent
{\bf\large Acknowledgements:}
We thank Hans-Walter Rix for stimulating discussions. This work was
supported by the ``Sonderforschungsbereich 375-95 f\"ur
Astro-Teilchenphysik'' der Deutschen Forschungsgemeinschaft.

\end{document}